\documentclass[twocolumn,times]{aastex62}
\received{XXXX XX, 201X}
\revised{XXXX XX, 201X}
\accepted{\today}
\submitjournal{ApJ}

\shorttitle{Departures from the BLR scaling in AGN}
\shortauthors{Czerny et al.}

\begin{document}

\title{\large Interpretation of Departure from the Broad Line Region Scaling in Active Galactic Nuclei}

\correspondingauthor{B. Czerny}
\email{bcz@cft.edu.pl}

\author{Bo{\.z}ena Czerny}
\affiliation{Center for Theoretical Physics, Polish Academy of Sciences, Al. Lotnikow 32/46, 02-668 Warsaw, Poland}

\author{Jian-Min Wang}
\affiliation{Key  Laboratory  for  Particle  Astrophysics,  Institute  of  High  Energy
  Physics,   Chinese  Academy  of  Sciences,   19B  Yuquan  Road,   Beijing 100049, China}

\author{Pu Du}
\affiliation{Key  Laboratory  for  Particle  Astrophysics,  Institute  of  High  Energy
  Physics,   Chinese  Academy  of  Sciences,   19B  Yuquan  Road,   Beijing 100049, China}

\author{Krzysztof Hryniewicz}
\affiliation{Nicolaus Copernicus Astronomical Center, Polish Academy of Sciences, ul. Bartycka 18, 00-716 Warsaw, Poland}

\author{Vladimir Karas}
\affiliation{Astronomical Institute, Academy of Sciences, Bocni II, CZ-141 31 Prague, Czech Republic}

\author{Yan-Rong Li}
\affiliation{Key  Laboratory  for  Particle  Astrophysics,  Institute  of  High  Energy
  Physics,   Chinese  Academy  of  Sciences,   19B  Yuquan  Road,   Beijing 100049, China}

\author{Swayamtrupta Panda}
\affiliation{Center for Theoretical Physics, Polish Academy of Sciences, Al. Lotnikow 32/46, 02-668 Warsaw, Poland}
\affiliation{Nicolaus Copernicus Astronomical Center (PAN), ul. Bartycka 18, 00-716 Warsaw, Poland}

\author{Marzena Sniegowska}
\affiliation{Center for Theoretical Physics, Polish Academy of Sciences, Al. Lotnikow 32/46, 02-668 Warsaw, Poland}
\affiliation{Warsaw University Observatory, Al. Ujazdowskie 4, 88-478 Warsaw, Poland}

\author{Conor Wildy}
\affiliation{Center for Theoretical Physics, Polish Academy of Sciences, Al. Lotnikow 32/46, 02-668 Warsaw, Poland}

\author{Ye-Fei Yuan}
\affiliation{Astronomy Department, University of Science and Technology of China, Hefei 230026, China}



\begin{abstract}
Most results of the reverberation monitoring of active galaxies showed a universal scaling 
of the time delay of the H$\beta$ emission region with the monochromatic flux at 5100 \AA, 
with very small dipersion. Such a scaling favored the dust-based formation mechanism of the 
Broad Line Region (BLR). Recent reverberation measurements showed that actually a significant 
fraction of objects exhibit shorter lags than the previously found scaling.
Here we demonstrate that 
these shorter largs can be explained by the old concept of scaling of the BLR 
size with the ionization parameter. Assuming a universal value of this parameter and 
universal value of the cloud density reproduces the distribution of observational points 
in the time delay-monochromatic flux plane, provided that a range of black hole spins is allowed. 
However, a confirmation of the new measurements for low/moderate Eddington ratio sources 
is strongly needed before the dust-based origin of the BLR can be excluded. 
\end{abstract}




\section{Introduction}
\label{sec:intro}

The most prominent features in type 1 Active Galactic Nuclei (AGNs) are broad emission 
lines present in their spectra (for a review, see e.g. \citealt{krolik1999}). The Broad Line Region (BLR) is 
unresolved with current instruments but the reverberation mapping (RM) of nearby active 
galaxies pioneered by \citet{liutyi1977} and done much more intensively since the
1990' \citep[e.g.][]{peterson2004,kaspi2000,bentz2013,du2018} allowed to measure the size 
of the BLR from the time delay between the variations of emission lines and the continuum. This in 
turn opened a way to measure the black hole mass by combining the radius 
from the time delay with the orbital velocity of the BLR 
clouds estimated  from the emission line width and assuming the Kepler law for that purpose.

Numerous RM campaigns 
revealed a very strong and tight correlation 
between the BLR size and the monochromatic luminosity at 5100 \AA~\citep{peterson2004,bentz2013}. 
With this scaling, supermassive black hole mass measurements became possible, based just on a single 
spectrum \citep[e.g.][]{vestergaard2006}. This, in turn, opened a way for cosmological applications, 
since after proper calibration the time delay measurement allows to determine the luminosity, and 
to use a generalized standard candle approach to obtain the cosmological 
parameters \citep{watson2011,haas2011,czerny2013,king2014}. 

The problem has started with the detection of some outliers from the radius-luminosity relation. 
First, outliers have been found among the highly super-Eddington sources \citep{du2015,du2016,du2018}, 
and their much shorter time delay than implied by the standard radius-luminosity relation \citep{bentz2013} 
could be interpreted as an effect of the self-shielding in the disk emission \citep{wang_shield2014}.

Recently, shorter than expected time delays also were measured in some low Eddington ratio 
sources \citep{grier2017,du2018}. It poses a question about the nature of the standard radius-luminosity 
relation and the physical reasons for the departures from this law. These shortened lags could be explained 
by retrograde accretion \citep{wang2014,du2018}. Actually, the averaged radius of the BLR depends on the 
ionizing spectral energy distributions and spatial distributions of BLR clouds. The increase of scatterers 
around the canonical $R-L$ relation indicates a lack of understanding of the ionizing sources and 
the BLR itself.

Most of the models assume that the BLR is quasi-spherical, the radial extension of the cloud formation 
is not specified \citep[e.g.][]{pancoast2014}, and the clouds are exposed to the nuclear emission, so 
the BLR emissivity should respond to the bolometric luminosity of the nucleus or, more precisely, to 
the available ionizing continuum. This is why the ionization parameter $U$ \citep[see e.g.][]{wandel1999} 
was used in most of the past BLR modelling. Some models specified the inner BLR radius on some physical 
grounds (e.g. disk local self-gravity, \citealt{wang2011,wang2012}; dust presence in the accretion disk 
atmosphere, \citealt{czkr2011,czerny2015,czerny2017}).

However, the relation between the 5100 \AA~ monochromatic luminosity, $L_{5100}$, and the ionizing flux 
is non-linear and depends on the Spectral Energy Distribution (SED) of the source. Large black hole mass, 
low Eddington ratio and low spin lead to significant curvature of the spectrum in the UV 
band \citep[e.g.][]{richards2006,capellupo2015}. In lower Eddington sources the inner disk may not be 
well represented by the standard accretion disk which effectively gives a similar 
result \citep[see e.g.][and the references therein]{kubota2018}. Therefore, as argued by \citet{wang2014}, 
we should expect significant dispersion if we use $L_{5100}$ as a proxy for the ionizing flux, unless the 
spectrum shows no curvature due to a very high black hole spin in all sources. The need to
 return to the scaling with bolometric luminosity has been suggested by \citet{trippe2015}. \citet{eser2015} showed 
observationally using seven well studied AGN that the relation between UV and optical flux is non-linear, 
and UV flux offers much better proxy for the ionizing flux and leads to the lower dispersion in the 
radius-luminosity relation.  

In the present paper we adopt a general approach based on the assumption that the size of the BLR scales 
with the ionizing flux. However, as in \citet{wang2014}, we take into account that the ionizing flux 
is not a linear function of the monochromatic luminosity. We consider the predictions for the ionizing 
flux from an accretion disk around a spinning black hole, together with the possibility of a 
counter-rotating disk and the possible existence of the inner hot flow instead of a standard optically 
thick cold accretion disk.

\section{Method}
\label{sec:method}

We consider the standard scenario of the BLR sensitive to the full ionizing flux available. In this model, 
the relation between the ionizing flux and the monochromatic luminosity at 5100 \AA~ is not linear, and 
the size of the BLR is expected to depend on the black hole mass and the Eddington ratio. Additionally, 
there are two effects: (i) the spin of the black hole plays an important role, (ii) in the case of low 
Eddington ratio AGN the inner radius of the optically thick
accretion disk might not be located at Innermost Stable 
Circular Orbit (ISCO) but further out, and in the innermost part the flow is replaced by optically thin 
and hot Advection Diminated Accretion Flow (ADAF). We consider these effects separately.
The role of the spin as a possible source of dispersion 
and systematic departures from the simple radius-luminosity trend has already been studied 
by \citet{wang2014}, but here we generalize the method and study the latter option of the inner ADAF.

\subsection{Accretion disk model and ionizing flux}

Since \citet{wang2014} showed that, most likely, a high value of spin is required,
in the present paper we use 
the Novikov-Thorne model of the accretion disk \citep{novikov1973,wang2014}. We combine the local 
emissivity with  the ray tracing procedure applied earlier in \citet{czerny2011} which allows us to 
include all relativistic corrections, including the light bending, gravitational redshift and Doppler 
boosting. The disk model is thus parametrized by the black hole mass, $M_{\bullet}$, accretion rate, 
$\dot M_{\bullet}$, and dimensionless spin parameter, $a$. We consider both prograde and retrograde 
spin values. 

The calculation of the observed monochromatic luminosity in this model depends on the viewing angle, 
$i$, between the symmetry axis and the observer. We do not know this angle in individual sources, 
but the expected range of angles is
constrained by the presence of the dusty-molecular torus \citep[see][]{krolik1999}. 
The torus blocks the view toward the nucleus at high viewing angles, and such sources do not show their BLR, 
and they are classified as type 2 sources. Observational constraints on the torus opening angle are not simple,
they typically imply a mean torus opening angle of order of $45^{\circ}$, and this value does not depend strongly on the source 
luminosity \citep[e.g.][]{tovmassian2001,lawrence2010,he2018}. Taking into account this constraint and 
assuming otherwise random orientation of AGN with respect to us we adopt the viewing angle $i \sim 30^{\circ}$ as 
a representative value. Thus we measure the observed fluxes assuming the same viewing angle for all the sources. 
We neglect here the effect of finite wavelength width for the hydrogen cross-section (see Eq. 5 
of \citealt{wang2014}) and determine the luminosity at 13.6 eV (1 Rydberg) as
\begin{equation}
  L_{\rm ion} = \nu L_{\nu} (1\,{\rm Rydberg})~ [{\rm erg~s}^{-1}].
\label{eq:L_ion}
\end{equation}

However, the BLR sees a different part of the spectrum. The BLR covers from 10\% to 30\% of the sky from the 
point of view of the inner disk, and it intercepts photons propagating relatively close to the equatorial 
plane (but not too close since highly inclined photons are intercepted by the disk itself). So, when 
calculating the number of ionizing photons we integrate the spectrum above 1 Ry over all viewing angles 
between $80^{\circ}$ and $45^{\circ}$ 
\begin{equation}
Q = \int_{45^{\circ}}^{80^{\circ}}\int_{1\,{\rm Rydberg}}^{\infty} {L_{\nu}(\theta) \over h \nu}d\nu \sin \theta d\theta ~[{\rm photons~s}^{-1}].
\label{eq:Q}
\end{equation} 
The integration is performed far from the black hole, where the GR effects are already negligible. 
Photons propagating at angles larger than $80^{\circ}$ are neglected since they will be absorbed by the disk 
itself. The ionization level of the BLR clouds can be estimated using either $L_{\rm ion}$, or $Q$.

\subsection{Accretion disks with inner ADAF}

The transition from the outer cold disk to the inner hot disk is still debated.  particularly for the 
galactic sources. However, it is clear that in very low luminosity sources (the most extreme case is Sgr A*) 
there is no cold outer disk. A series of papers discussed this issue on the basis of the radiative and 
conductive interaction between the hot corona above the disk and the underlying cold disk, and it was 
shown that for very low accretion rates the cold inner disk disappears \citep[see][for a review]{yuan2014}. 
The inner flow then proceeds through an optically thin hot flow, such as ADAF \citep[][]{ichimaru1977,narayan1994} or its alternatives, e.g. Advection Dominated Inflow-Outflow 
Solutions \citep[ADIOS;][]{blandford1999}. Most of these solutions can be described under the common name 
RIAF (Radiatively Inefficient Accretion Flow) but some are actually quite radiatively efficient if the 
strong coupling exists between the hot ions and electrons \citep{bisnovatyi1997,sironi2015}. However, 
the common property of these solutions is that ion temperature is close to virial temperature, and 
electrons are also relatively hot so the emitted radiation concentrates in X-rays instead of far-UV local 
black body emission characteristic for the cold accretion disks.

For the purpose of this paper we use 
two prescriptions from \citet{czerny2004}. The first one is simply based on the {\it strong ADAF principle}, 
i.e. whenever the ADAF solutions exists the flow proceeds through an ADAF-type flow, as in the classical papers \citep{abramowicz1995,honma1996,kato1998}. The second option is  
based on evaporation of the cold disk caused by electron conduction between the disk and the two-temperature 
hot corona \citep{rozanska00,meyer02}.

In the first case the transition from a cold disk to an ADAF flow occurs at 
\begin{equation}
  R_{\rm evap} = 2.0\, \alpha_{0.1}^{4}\dot m^{-2} R_{\rm Schw},  
\label{eq:ADAF}
\end{equation}
(see e.g.~Eq. 8 in \citealt{czerny2004}), where $\dot m$ is the dimensionless 
accretion rate, $\alpha_{0.1}$ is the viscosity parameter in units of 0.1, and $R_{\rm Schw}$ is the  Schwarzschild 
radius of the black hole (equal 2 $R_{\rm g} = 2 GM/c^2$, ). Here $\dot{m}$ is
defined for a fixed Newtonian efficiency of the accretion process
\begin{equation}
  \dot m = {\dot M_{\bullet} \over \dot M_{\rm Edd}}; ~~~~ 
  \dot M_{\rm Edd} = {48\pi G M_{\bullet} m_{\rm p} \over \sigma_{\rm T} c},
\end{equation}
where $m_{\rm p}$ is the proton mass, and $\sigma_{\rm T}$ is the Thomson cross-section.

In the description of the second scenario we adopt Eq.~11 from \citet{czerny2004} 
for the transition radius between the outer disk and an inner hot flow since it contains the effect of the 
magnetic pressure and compares most favourably with the observed extension of the BLR:
\begin{equation}
  R_{\rm evap} = 19.5\, \alpha_{0.1}^{0.8}\beta^{-1.08}\dot m^{-0.53} R_{\rm Schw},
\label{eq:evap}
\end{equation} 
where $\beta$ is the ratio of the total (gas + radiation) pressure to the total plus magnetic pressure and 
varies between 0 (magnetic pressure dominance) to 1 (no magnetic pressure). 
We neglect the emission from the inner hot flow because this very hot plasma, with electron temperature of 
the order of tens of keV, does not contribute to the 5100 \AA~ flux. This emission can to some extent affect 
the BLR by Compton-heating the clouds and the intercloud medium suppressing the line emission. However, 
since we are interested only in the ionization flux, and not in full radiative transfer computations with 
cooling/heating balance, we neglect this emission, effectively locating the inner disk radius at 
$R_{\rm evap}$.
In this scenario the dependence on the spin practically disappears since the outer disk is only weakly 
affected by the rotation of the central black hole. 

\subsection{BLR radius}

We assume the standard view that the localization of the BLR is related to the ionizing flux. When we use 
$L_{\rm ion}$ defined in Eq.~\ref{eq:L_ion} we follow the approach of \citet{wang2014}. We assume the scaling 
of $R_{\rm BLR}$ with the ionizing flux to have the form of a power law with the same index as determined 
by \cite{bentz2013},
\begin{equation}
  \log R_{\rm BLR} = 0.542 \log L_{\rm ion} + const,
\end{equation}
where we specifically took the value of the index from their fit {\it Clean}. The value of constant has 
to be adjusted since now we use $L_{\rm ion}$ instead of $L_{5100}$ as done in \cite{bentz2013}.

When we use $Q$ as the parametrization of the incident flux, we follow even more classical approach to 
modelling of the BLR. It was argued in the past that the BLR properties are
well approximated by the fixed value of the ionization parameter, $U$, and the representative cloud density.
The ionization parameter $U$ is defined as
\begin{equation}
U = {Q \over 4 \pi r^2 c n_e},
\end{equation}
\citep[see e.g.][]{ferland83} where $n_e$ is the representative local density of the cloud, and $Q$ is the 
number of ionizing photons (above 1 Rydberg) emitted by the accretion disk. We thus calculate the BLR radius 
from this formula,
\begin{equation}
\log R_{\rm BLR} = 0.5 \log Q + const. 
\end{equation}
The value of the constant is then related to the universal values of $U$ and $n_e$. The universal characteristic 
of the cloud density was argued for at the basis of the radiation pressure confinement mechanism \citep{baskin2018}. 

\subsection{Observational data}
\label{sect:obsdata}

We compare the model with the size of the BLR measured as a delay with respect to H$\beta$ line in reverberation 
campaigns. We use a compilation of the results available in the literature (see Table~\ref{tab:objects}). The 
measurements come from various groups, most of them were performed for nearby sources. The sample 
of \citet{bentz2013} has been carefully corrected for the contamination of the 5100 \AA~flux by the host galaxy. 
The sample from \citet{grier2017} comes from SDSS-RM (Sloan Digital Sky Survey Reverberation Measurement) project 
and covers larger redshifts up to $z  = 1.026$. The measurements by \citet{lu2016} provide an independent 
determination of the delay in NGC 5548, and \citet{wang2016} gives the delay measured for a gamma-ray loud NLS1.  
The sample from \citet{du2014,du2015,du2016,du2018} papers represent SEAMBH (Super-Eddington Accretion in Massive 
Black Holes) project so on average these objects have higher Eddington ratios than sources from other samples. 
In diagrams we mark them with a different colour as they might bias the results. We also give in Table~\ref{tab:objects}
the black hole mass values taken from the references above, and the Eddington ratio which we calculate from that mass value and from  the monochromatic luminosity assuming a fixed bolometric correction of 9.26 after \citet{shen2011}. Absolute values of the luminosity
are given assuming the cosmological parameters: $H_0=67$ km s$^{-1}$ Mpc$^{-1}$, $\Omega_{\Lambda}=0.68$, $\Omega_m=0.32$ \citep{planck2013}.

\section{Results}
\label{Results}

We compare the measured time delays in monitored AGN with the model prediction of the position 
of the BLR radius taking into account two ways of connecting the BLR position with the incident 
radiation (through $L_{\rm ion}$ defined in Eq.~\ref{eq:L_ion} and $Q$ defined in Eq.~\ref{eq:Q}). 
The incident spectrum is calculated for a realistic range of black hole masses 
(from $10^6 M_{\odot}$ to $10^{10} M_{\odot}$) and Eddington luminosities (from 0.01 to 1.0). 
We allow for a broad range of spin, including the case of a counter-rotating black hole, and we also allow 
for the evaporation of the inner disk given by Eq.~\ref{eq:evap}.

\begin{figure*}
\centering
\includegraphics[width=0.95\textwidth]{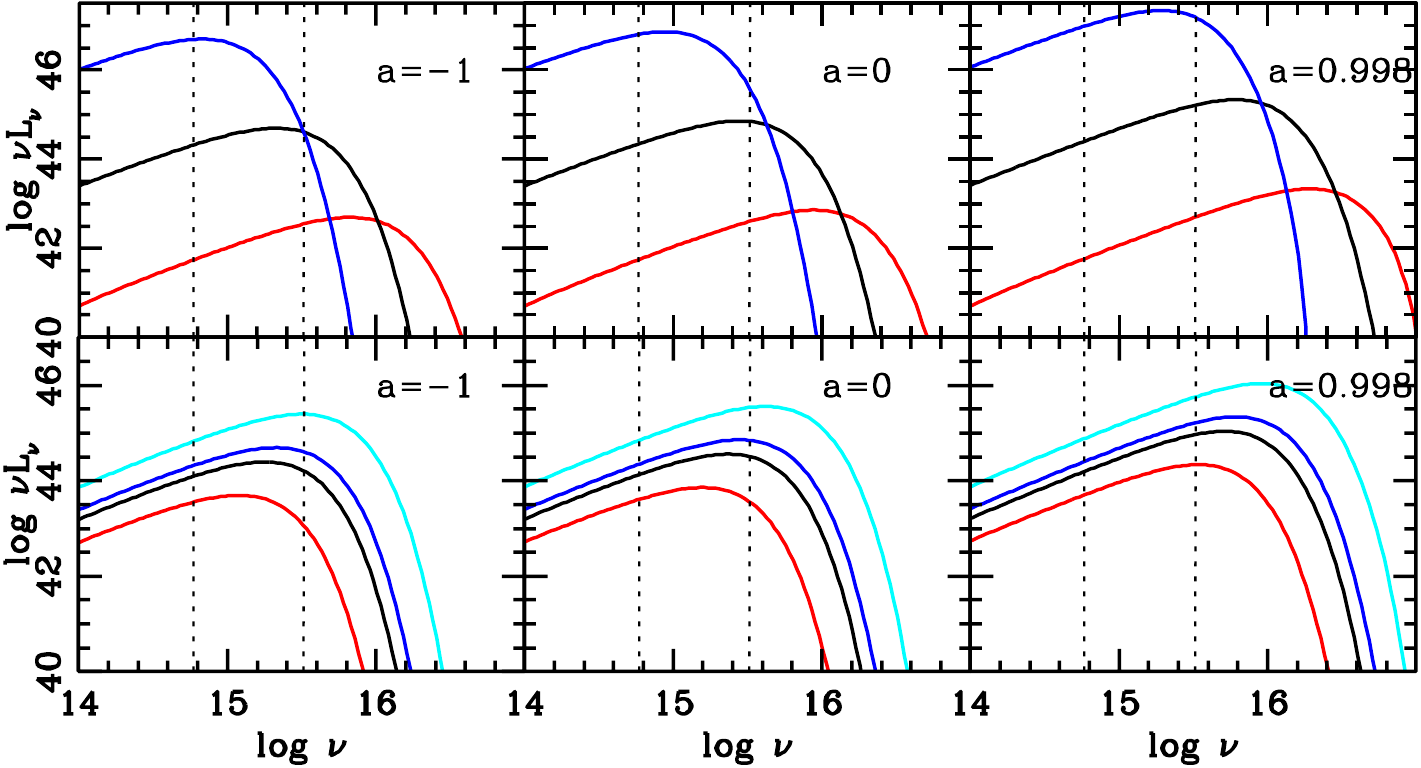}
\caption{The examples of the accretion disk spectra for the parameter range covered by our 
computations. Upper panels shows spectra for a fixed Eddington ratio 0.1, and black hole 
masses $10^6 M_{\odot}$ (red), $10^8 M_{\odot}$ (black), and $10^{10} M_{\odot}$ (blue lines), lower panel shows spectra 
for a black hole mass $10^8 M_{\odot}$ and eddington ratio 0.01 (red), 0.05 (black), 0.1 (blue), and 0.5 (cyan). The three columns 
represent three spin values. Dotted lines indicate the position of the usual continuum 
measurement at 5100 \AA, and the Lyman edge where ionizing flux should be evaluated.}
 \label{fig:spectra}
\end{figure*}

Since the observational results are always plotted as a delay versus the monochromatic flux at 
5100 \AA~ but the ionization flux in general  is not a linear function of that flux, we first 
show the representative examples of the incident spectra for the adopted parameter range (see 
Fig.~\ref{fig:spectra}). The relation between $L_{\rm ion}$ and the 5100 \AA~ luminosity is almost 
linear when the Eddington ratio is large, the black hole mass is small, and the black hole spin  
large: in this regime the spectrum up to 912 \AA~ is still well described by the canonical 
$ \nu L_{\nu} \propto \nu^{4/3}$ law. Outside this regime, the maximum temperature of the accretion 
disk drops, and the SED peak moves into UV band, leading to a strong spectral curvature. A similar 
effect would be caused by the inner disk evaporation but the presented examples illustrate 
accretion disks extending down to Innermost Stable Circular Orbit (ISCO). The disk spectral curvature leads to a slower rise of 
$L_{\rm ion}$ in comparison to $L_{5100}$ in a sequence of models with a constant Eddington rate and 
rising black hole mass. We take into account this effect in interpreting the data, since we 
calculate the expected time delay from $L_{\rm ion}$ (or $Q$).

\begin{figure*}
\centering
\includegraphics[width=0.95\textwidth]{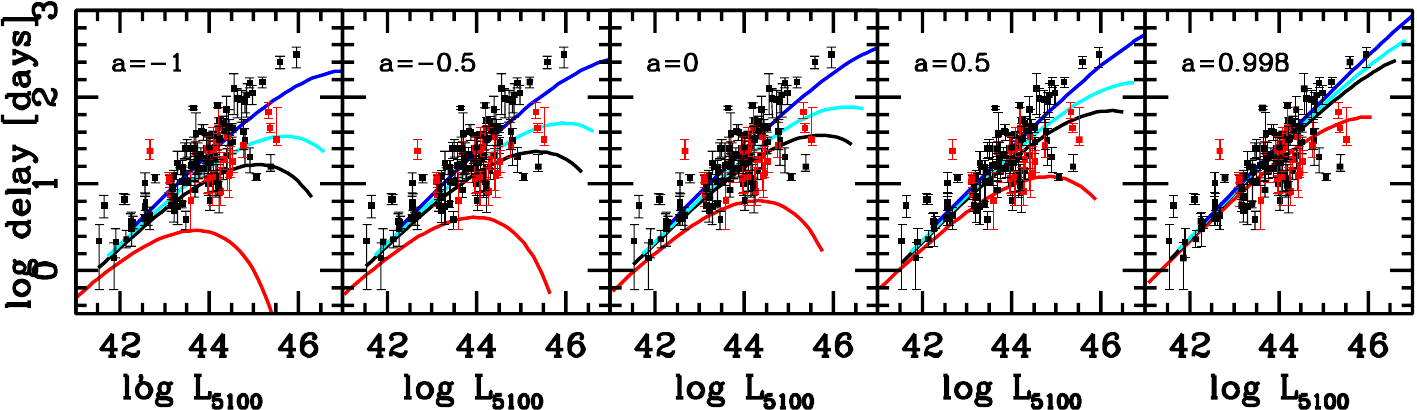}
\caption{The dependence of the size of the BLR, expressed in terms of the time delay, calculated from $L_{\rm ion}$ for a range of spin
(separate panels), and Eddington ratio $\dot m = 0.01$ (red line), $\dot m = 0.05$ (black line), 
$\dot m = 0.1$ (cyan line), and $\dot m = 0.5$ (blue line) as a function of monochromatic luminosity 
at 5100 \AA. Observational points come from Table~\ref{tab:objects}, super-Eddington sources are 
marked in red.}
 \label{fig:no_evap}
\end{figure*}

We now calculate the expected time delays from our grid of models at the basis of the ionized flux, 
but we show the resulting delays as a function of the monochromatic flux since the observational 
results are customarily presented in this way. We compare the predicted delay grid with the measurements of the 
H$\beta$ delays given in Table~\ref{tab:objects}.

\subsection{BLR size from the ionizing luminosity}

Since our sample is larger than the sample considered in \citet{wang2014}, we first use the same 
method as was used there, with BLR distance measured from ionizing flux  $L_{\rm ion}$ estimated at 
912 \AA ~(see Eq.~\ref{eq:L_ion}), and without any evaporation effect. The results are shown in 
Fig.~\ref{fig:no_evap}.

\begin{figure*}
\centering
\includegraphics[width=0.95\textwidth]{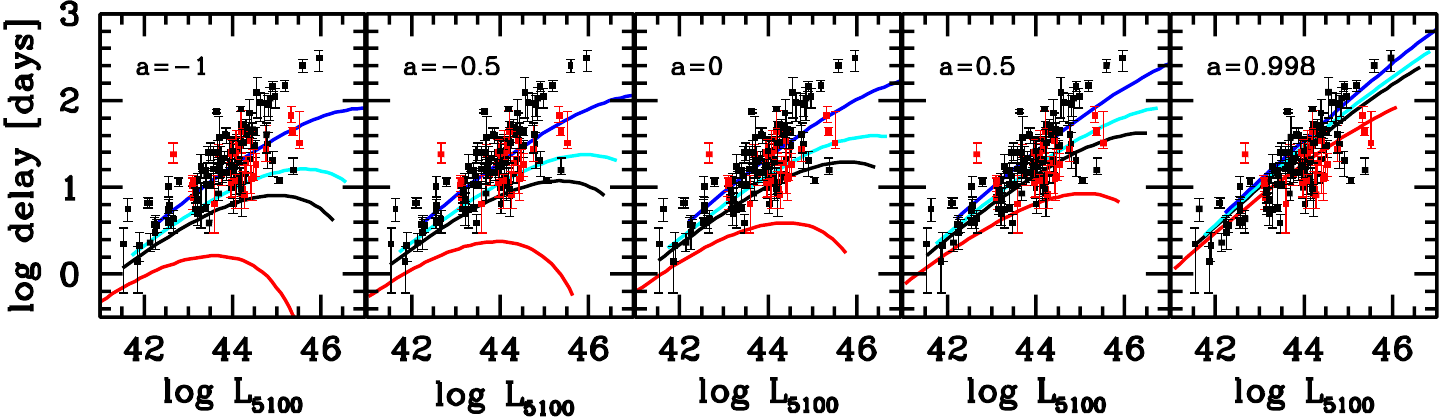}
\caption{The same as Fig.~\ref{fig:no_evap} but for the size of the BLR calculated from $Q$.}
 \label{fig:no_evap_Q}
\end{figure*}

We see that the change of the sample essentially affects the conclusion reached by \citet{wang2014}. 
Their conclusion was that all the monitored objects must have predominantly high spin since at that 
time most of the measured delays were located along the $\tau \propto L_{5100}^{1/2}$ line. Currently, 
with the presence of many shorter lags in the sample, good coverage of the measured distribution of 
the time delays is achieved if we allow a whole range of black hole spins from 0 to maximally rotating 
black holes. A few objects require counter-rotating spin. What is interesting, however, these objects 
have rather moderate luminosities, $\sim 10^{44}$ erg s$^{-1}$, relatively low Eddington ratios, 
$\sim 0.05$, and black hole masses which are also moderate, $\sim 10^{8} - 10^{9} M_{\odot}$. These 
values are roughly consistent with the parameters of the sources in the SDSS-RM sample of \citet{grier2017}. 
Such accretion rates are relatively low for a quasar sample but this is the consequence of the selection effect: 
the monitoring was short, about a year, so the delays were measured only for the sources with delays shorter 
than 100 days in the observed frame, so only for the low luminosity tail of the sample. The model predicts 
even shorter time delays than measured for more massive sources but observationally determined delays do 
not populate this region. This is because we allowed for even lower Eddington ratios (0.01) and larger 
masses ($10^{10} M_{\odot}$) in our parameter grid than present in the sample.

\begin{figure}
\centering
\includegraphics[width=0.45\textwidth]{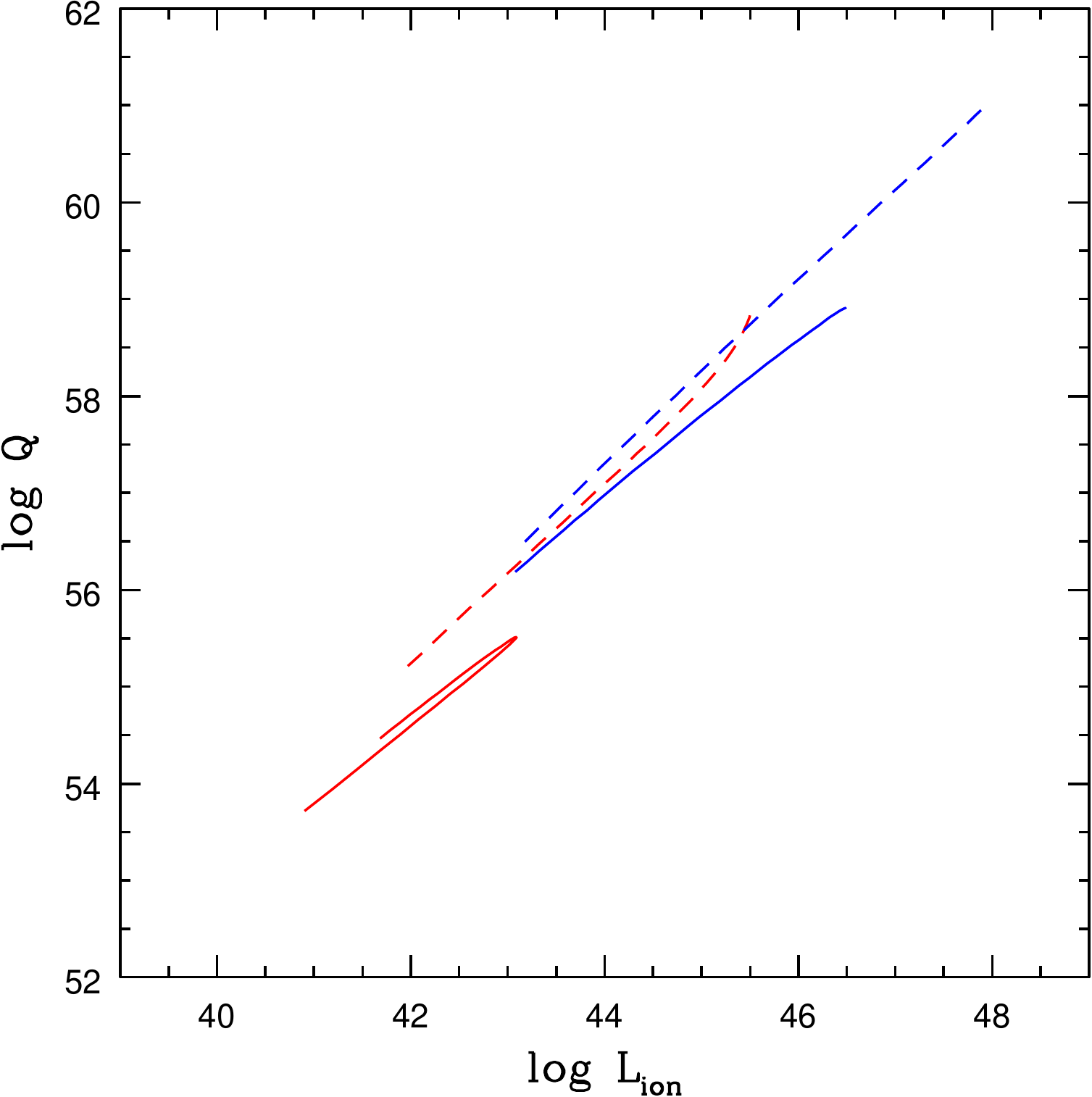}
\caption{The relation between the ionization luminosity $L_{\rm ion}$ and the number of ionizing photons, 
$Q$, for the size of the BLR for $a = -1.0$ (dashed lines), and $a = 0.998$ (continuous lines), for 
two values of the accretion rate $\dot m = 0.01$ (red lines), and $\dot m = 0.5$ (blue lines). Both 
$L_{\rm ion}$ and $Q$ are not monotonic functions of the black hole mass, as reflected on turn-over loop,
but the two quantities are not strictly proportional to each other.}
 \label{fig:Lion_Q}
\end{figure}

\subsection{BLR size from the number of ionizing photons}

Next we use the $R_{\rm BLR}$ predictions based on the number of photons $Q$ intercepted by the BLR (see 
Eq.~\ref{eq:Q}), again without any evaporation effect, and the results are shown in Fig.~\ref{fig:no_evap_Q}. 
The model predictions are qualitatively similar, but not identical. Overall, $Q$ prescription gives much 
stronger bending effect for a counter-rotating accretion disk than the model based on $L_{\rm ion}$ for the 
same parameters. The trend reverses for high spin co-rotating accretion disk, when smaller departure 
from a power law trend is seen for $Q$-based model. This is related to the relativistic effects. For large 
black hole spin in a corotating disk the radiation emitted at higher inclination angles is strongly 
beamed and enhanced in comparison with the continuum measured by an observer, which compensates for 
the spectral bending seen in Fig.~\ref{fig:spectra}. This also means that $Q$ is not strictly proportional 
to $L_{\rm ion}$. We illustrate this effect in Fig.~\ref{fig:Lion_Q}. The departure from the strict linearity 
between the two quantities results both from the derivation of $Q$ as an integral, and from the relativistic 
effects (a difference between photons going to observer and photons going toward BLR). The approach based 
on $Q$ is more accurate, and the examples calculated later on are based on this assumption but overall 
the difference between $L_{\rm ion}$ and $Q$ predictions is not large and $L_{\rm ion}$ approach is equally 
useful for statistical analysis of the samples.

Now we compare $Q$-based delay predictions to the measured time delays (see Fig.~\ref{fig:no_evap_Q}). 
This approach does not require counter-rotating black holes to explain even the shortest time lags. 
Therefore, with very simple and basic assumptions (standard Novikov-Thorne disk, co-rotating with a 
black hole, with a range of masses, accretion rates and spins, BLR responding to the ionizing continuum 
at fixed ionization parameter and density) we are able to reproduce fully satisfactorily the observed 
distribution of the time delays.

\begin{figure*}
\centering
\includegraphics[width=0.95\textwidth]{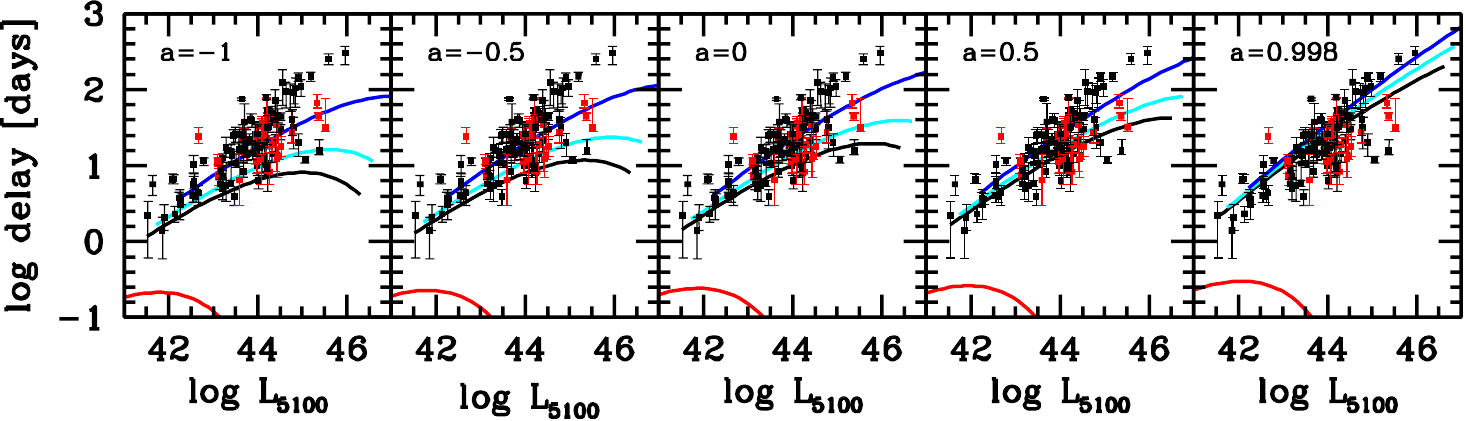}
\caption{The dependence of the size of the BLR calculated from $Q$ when ADAF principle is assumed, 
for a range of spins (separate panels), and Eddington ratios $\dot m = 0.01$ (red line), $\dot m = 0.05$ 
(black line), $\dot m = 0.1$ (cyan line), and $\dot m = 0.5$ (blue line) as a function of monochromatic 
luminosity at 5100 \AA. Model parameter: $\alpha = 0.02$. Observational points come from 
Table~\ref{tab:objects}, super-Eddington sources are marked in red.}
 \label{fig:evap_Q_ADAF}
\end{figure*}

However, the assumption that the cold Keplerian disk extends all the way down to ISCO is under 
discussion, particularly for lower Eddington ratio sources. If the inner hot flow develops, 
this part of the disk is not longer a source of UV photons but instead, filled with a hot plasma 
at electron temperatures of order of 100 keV is a source of X-ray emission. Effectively this 
decreases the number of photons available for ionizing hydrogen. Available models allow to 
determine the position of this transition and thus are subject of tests if the predictions 
are consistent with the observed time delays. 

\subsection{BLR size with inner hot flow in classical ADAF scenario}

We first test the prediction of the model based on {\it strong ADAF principle} described by 
Eq.~\ref{eq:ADAF}. For the viscosity parameter, we assume the value $\alpha = 0.02$ which 
was suggested by direct studies of the quasar UV variability \citep{siemiginowska1989,starling2004}, 
and is consistent with Damped Random Walk results (\citealt{kelly2009}; see also the discussion 
in \citealt{grzedzielski2017}). The results are shown in Fig.~\ref{fig:evap_Q_ADAF}. In this 
case no counter-rotating spin values are necessary, and the whole plane is well covered even 
if all the objects have high spin (left panel). The longest delays for a given value of $L_{5100}$ 
require high spin values while shorter delays may imply either lower spin or lower Eddington ratio. 
This degeneracy can be removed if we actually have reliable determination of the Eddington ratio 
for an individual object. However, in comparison with the predictions for the disk without an 
inner cut-off the requested accretion rates of sources with short delays are much higher, all 
observed sources are then predicted to have Eddington ratios above $\sim 0.05$. Some of the 
Eddington ratios in the SDSS-RM sample may be lower than this limit if the bolometric luminosity 
is estimated as roughly 9 times the $L_{5100}$ luminosity and the black hole mass measurement 
from \citet{grier2017} is adopted. On the other hand, such an estimate of the bolometric luminosity 
is highly uncertain since it does not take into account the differences in the spectral shapes 
clearly seen in Fig.~\ref{fig:spectra}.

\begin{figure}
\centering
\includegraphics[width=0.45\textwidth]{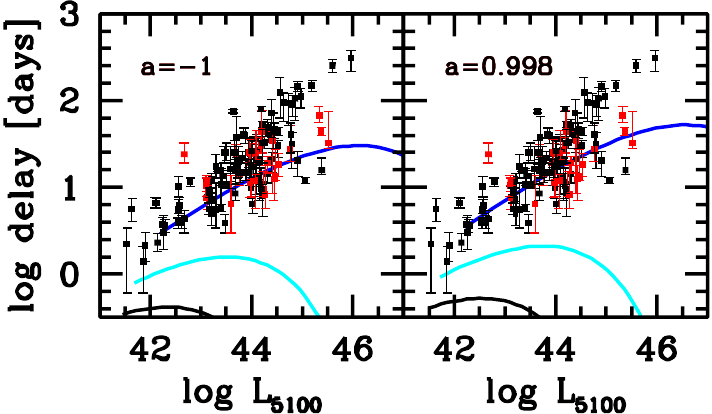}
\caption{The dependence of the size of the BLR calculated from $Q$ when evaporation of the 
inner disk is included, for a maximally counter-rotating spin (left panel, ISCO radius 9 $R_g$), and maximally 
co-rotating spin (right panel, ISCO radius 1.24 $R_g$), and Eddington ratios $\dot m = 0.01$ (red line), $\dot m = 0.05$ 
(black line), $\dot m = 0.1$ (cyan line), and $\dot m = 0.5$ (blue line) as a function of 
monochromatic luminosity at 5100 \AA. Evaporation model parameters: $\alpha = 0.02$,  
$\beta = 0.5$. Observational points come from Table~\ref{tab:objects}, super-Eddington 
sources are marked in red.}
 \label{fig:evap_Q_0.5}
\end{figure}

\begin{figure}
\centering
\includegraphics[width=0.41\textwidth]{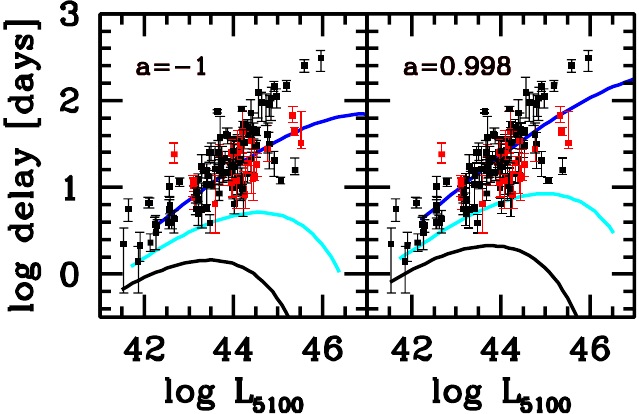}
\caption{The same as Fig.~\ref{fig:evap_Q_0.5} but for the magnetization parameter $\beta = 0.99$.}
 \label{fig:evap_Q_0.99}
\end{figure}

\begin{figure*}
\centering
\includegraphics[width=0.95\textwidth]{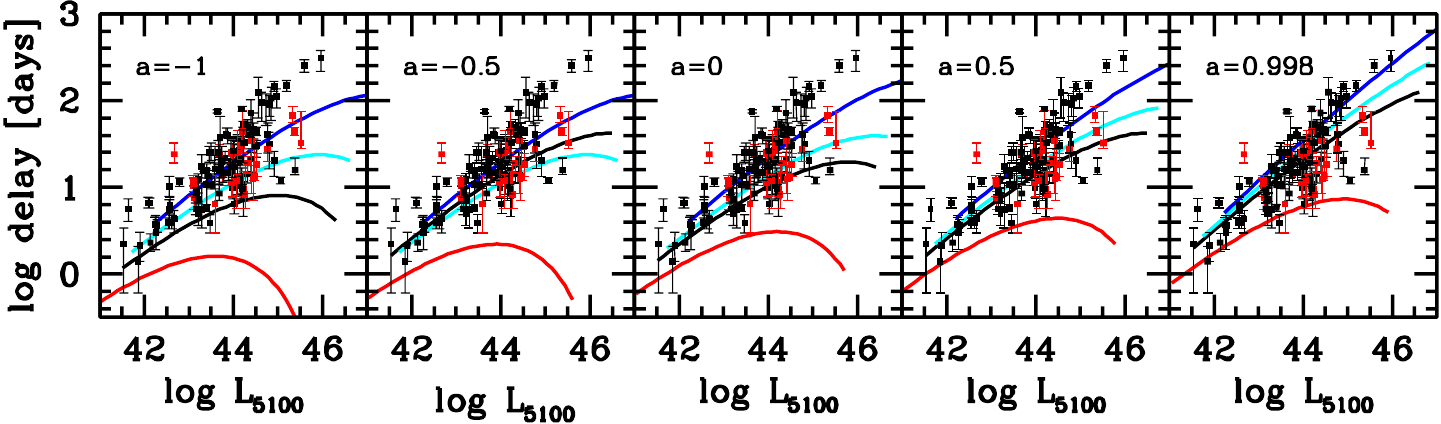}
\caption{The dependence of the size of the BLR calculated from $Q$ when evaporation of the 
inner disk is included, for a range of spin (separate panels), and Eddington ratio 
$\dot m = 0.01$ (red line), $\dot m = 0.05$ (black line), $\dot m = 0.1$ (cyan line), and 
$\dot m = 0.5$ (blue line) as a function of monochromatic luminosity at 5100 \AA. Evaporation 
model parameters: $\alpha = 0.001$,  $\beta = 0.99$. Observational points come from 
Table~\ref{tab:objects}, super-Eddington sources are marked in red.}
 \label{fig:evap_Q_alpha_mala_0.99}
\end{figure*}

\subsection{BLR size with inner hot flow defined by inner disk evaporation}

Finally, we test the solutions based on the evaporation of the inner disk as described by 
Eq.~\ref{eq:evap}. The results obtained for the parameters $\alpha = 0.02$ and $\beta = 0.5$ 
(moderate magnetic field strength; see Eq.~\ref{eq:evap}) are far from being satisfactory 
(see Fig.~\ref{fig:evap_Q_0.5}). The longest time delays are not reproduced, even if we allow 
for high spin since the transition radius between the outer cold disk and an inner hot flow 
is far too large. Significant contribution of the magnetic field to the total pressure in 
the disk is inconsistent with the observational data. We thus decreased the role of the 
magnetic field, assuming $\beta = 0.99$ (i.e. only 1 \% of the contribution from the magnetic 
pressure to the gas plus radiation pressure) but such parameter adjustment still did not fully 
solve the problem (see Fig.~\ref{fig:evap_Q_0.99}). We thus additionally allowed for a 
significant decrease of the viscosity parameter down to the value $\alpha = 0.001$. For these 
parameters, evaporation of the outer cold disk is indeed inefficient. The transition radius 
between the outer cold disk and an inner hot flow takes place at $5.7 R_{\rm Schw}$ for the 
Eddington ratio 0.01 and it is closer in or absent for larger accretion rates and/or lower 
spin. When we assumed the expected values of the viscosity parameter $\alpha = 0.02$, the 
transition radii start at $62.3 R_{\rm Schw}$ (for the lowest Eddington ratio 0.01). This low 
viscosity $\alpha \sim 0.001$ is thus strongly required if the measured time delays are to 
be consistent with the disk evaporation phenomenon. 

However, as we mentioned before, such 
small values of the viscosity parameter are not supported by the quasar variability 
(see \citealt{grzedzielski2017}). This implies that the physically-based description of 
the disk evaporation and the transition to inner hot flow still requires a more advanced 
approach. The issue is difficult, and subject of studies since several years in the context 
of cataclysmic variables \citep{meyer1994}, binary black holes and AGN \citep{liu1999,rozanska2000}. 
The transition process is further complicated by the possibility of the existence of the 
inner cool disk separated by the gap of pure hot flow from the outer cold 
disk (\citealt{liu2006,meyer2007}; see \citealt{meyer2014,MeyerHofmeister2017,taam2018} 
for recent developments). Here we considered only the inner radius of the outer cold disk. 

\section{Discussion} \label{sec:discussion}
\subsection{The BLR models}
Emission lines of the BLR respond to the emission originating close to the black hole, as clearly 
seen from the response of the line with respect to the variable incident flux. Traditionally,
this response was modeled by parametrizing the incident continuum with a single parameter in the 
form of the ionization parameter, or ionizing flux. Then, if a range of radii and densities were 
allowed, like for example in the LOC model \citep{baldwin1995} line ratios were successfully 
modelled. However, this general approach did not give a direct insight into the reason for the 
presence of the numerous clouds above the disk. Wind models are also parametric and not quite 
constraining the cloud formation mechanisms. The observational discovery of the scaling of the 
BLR size with a monochromatic flux instead of the ionizing flux opened a way for testing the 
cloud origin. Such a scaling, with a very small dispersion, is consistent with the dust-based 
formation mechanism of the BLR \citep{czkr2011,czerny2015,czerny2017,baskin2018}, at least 
with respect to the H$\beta$ formation region. In this model the BLR position is controlled not by
the ionizing flux but by
the availability of the material for the irradiation, lifted above the disk by the
dust-driven radiation pressure. This scaling also strongly disfavoured self-gravitational 
instability as the BLR origin \citep{czerny2016}. 

New observational data presented in Fig.~\ref{fig:no_evap} drastically changes this view. Now the 
dispersion around the delay-monochromatic flux relation is very large, and most of the new 
measurements lie below the previous tight scaling law of \citet{bentz2013}. Part of the outliers 
come from the super-Eddington ratio sample by \citet{du2014,du2015,du2016,du2018}, and in this 
case the departure from the previous scaling can be explained as a departure from the thin 
Keplerian disk approximation for the incident continuum. The average value of the Eddingtn ratio in these sources is 1.72. However, other numerous points come 
from the SDSS-RM sample of \citet{grier2017}, and these objects have low to moderate Eddington ratios, with the average value of 0.24, not significantly higher than in the \citet{bentz2013} sample (0.14), well within the dispersion in the two samples.

  The measurements we use here come from the literature, as described in Sect.~\ref{sect:obsdata},
  and were not prepared in a uniform way. All sources in \citet{bentz2013}
  sample were carefully corrected for the host galaxy contamination using Hubble Space Telescope (HST) observations,
  and the
  corrections were important not just for the faintest nearby AGN but also for PG quasars \citep{bentz2006}. Good
  example from \citet{bentz2013} is the source SBS 116+583A ($z =  0.02787$) where the AGN emission contributes only
  12 \% to the total flux at 5100 \AA, i.e. not correcting this source for starlight gives a shift of 0.92 in the
  logarithm of $L_{5100}$, or, equivalently, a factor 2.9 in the expected time delay. The sources from
  \citet{du2014} were also corrected for the starlight using HST images, while for the sources reported later
  \citep{du2015,du2016,du2018} the authors used an empirical relation from \citet{shen2011}.  \citet{grier2017} used
  the information on host contamination from \citet{shen2015} who has employed Principal Component Analysis (PCA) and the measurements of the measured stellar velocity dispersion to decompose
  the spectra into AGN and host components. The accuracy of such methods is difficult to estimate. The error in the host subtraction
  can be responsible for the incorrect determination of $L_{5100}$. 

New results, if reliable, take us all the way back to the original concept of the simple BLR response to 
the ionizing flux. The range of delays are well consistent with the predictions of the ionized 
continuum based on the thin Keplerian disk since this ionized continuum does not scale linearly 
with the monochromatic flux at 5100 \AA ~ due to the curvature of the spectrum in the UV band (see 
Fig.~\ref{fig:spectra}). Models based on the dusty origin of the BLR seem not justified, as in this case 
no departure from \citet{bentz2013} scaling is predicted for a broad range of black hole mass 
and accretion rates.  Instead, the observed delays are consistent with a universal ionization 
parameter, and universal density in the H$\beta$ formation region for a very broad range of 
parameters.

The range of spin plays an important role, and all spins from non-rotating black 
holes to maximally spinning ones ($a = 0.998$) are requested to create the appropriate 
representation of the reverberation studies sources. Counter-rotating disks are not strongly 
required but a small fraction of such sources is not excluded. Evaporation of the inner disk 
and a transition to ADAF are not really required for Eddington ratios above 1\% studied here, 
and models which predict inner ADAF flow at such high Eddington ratios are disfavoured, so 
the models with an inner cold disk separated from the other cold disk sound more attractive. 
However, the predictions for such models are more complex and were not tested in detail in 
the current paper. In this paper we also did not test again the self-gravity scenario, but 
since self-gravity in general predicts shorter delays than \citet{bentz2013} relation it 
remains to check if this option offers equally good coverage of the parameter space as a 
simple universal ionization parameter with universal density model.

Recently, a new model of the BLR origin was suggested by \cite{wang2017}. In this model,
clumps in dusty/molecular torus are tidally captured and disrupted by the central black hole.
A population of inflowing 
clouds forms at the first stage, a tiny fraction of clumps is channelled 
into outflows (less than 10\%), and then most of the disrupted clumps form a BLR disk with Keplerian 
rotation (virialized component). Unlike the dust-based clouds \citep{czkr2011}, the supply of the 
BLR clouds originates from the dusty torus. The virialized component 
in this model can produce the canonical $R-L$ 
relation for sub-Eddington AGNs provided the ionisation parameter, density and temperature are
universal. Additionally, the infalling component is supported by PG 2130+099 (see Sect.~\ref{sect:future}). 
This model avoids difficulties of disk self-gravity model or dust-based failed winds.

\subsection{Implications to BH evolution}
What we found in this paper is the role of black hole spin in the observed $R-L$ relation
jointly with accretion rates.
The vindence in support of retrograde accretion onto black holes has very important implications for cosmological
evolution of black holes. If the black holes are fuelled in a stochastic manner, with no preferred orientation, they are slowly spinning due to
cancellation of random angular momentum of accreted gas \citep{king2008}. 
\cite{wang2009} built up an equation of the radiative efficiency ($\eta$) across cosmic time from observed data of galaxy and quasar
surveys. The authors derived the evolutionary curve and obtained that the radiative efficiency changes from $\eta\approx 0.3$ at redshift $z\approx 2$, where quasar density peaks, down to $0.03$ at
low redshift.  This supports the role of episodic accretion at later stages of the galaxy evolution.  The downsizing behaviour of spins was further discussed by \citet{Li2012}. Subsequently, \cite{volonteri2013} found a similar 
behaviours of spin evolution from numerical simulations. The sensitive dependence of the $R-L$ relation on
spin offers a new tool of estimating black hole rotation, in particular for those AGN with the retrograde 
accretion disks which may have too weak gravitational effects on iron K$\alpha$ line profiles to measure
their spins from  X-ray observations.  

\subsection{Future RM-campaigns}
\label{sect:future}
To draw firm conclusions, however, the measured delays have to be accurate.
SDSS-RM campaign concentrated on average on higher redshift sources, the campaign 
was relatively short, and the observational cadence was not very dense. Determination of 
the time delays in AGN is not very straightforward since the variability has a red noise character, 
and the resprocessing region is extended. Frequently, two peaks show up in the cross 
correlation function \citep[e.g.][]{du2016}, and if the campaign is too short only one 
solution (the shorter one) may be found although it may not be actually the correct one. 
Therefore, an extension of this campaign is clearly necessary to ensure that the measured 
delays are not affected by the way how they are performed.  

The cadence selected for RM-campaign is very important. Low-candence campaign will smear
short timescale variations so that the measured lags tend to be longer.
PG 2130+099 is an example. \cite{kaspi2000} measured a H$\beta$ line lag of 188\,days with 
a low cadence of about $\sim 20$ days and a couple of seasonal gaps, with the total campaign duration as long as 
about 8 years. With a cadence of a few days, however, \cite{grier2008} measured a lag 
of $\sim 23\,$ days with large uncertainties (but the total length in their case was only about 100 days). 
Moreover, \cite{grier2012} got a lag of $\sim10\,$days with one day cadence. Hu et al. (2018) (ApJ, submitted) 
measured a lag of $\sim 24\,$days with a cadence of 3\,days, 
confirming the results of \cite{grier2008}, but found that the lag of $\sim 188\,$days follows the 
dust reverberation scaling relation \citep{koshida2014}, suggesting that the reverberations with a 
lag of $188$ days
are from the inner edge of the torus. This supports the ideas of \cite{wang2017}. Additionally, 
PG 2130+099 is a super-Eddington source and the other two lags of $\sim 10$ and $\sim 24\,$days can be explained by the
self-shadowing effects of slim accretion disks (Wang et al. 2014b). Overall, the BLR is extended, and
the choice of the cadence can focus the monitoring on a particular part of the BLR, making a
comparison of results for different sources very difficult. Also non-linear response of the BLR to the line emission,
particularly if the characteristic variability timescale is short in comparison to the average time delay can easily lead to apparent shortening of the time delay, as pointed out by \citet{goad2014}. Future RM-campaigns
should be planned very carefully with respect to the cadence. 

\section{Conclusions}
\label{sec:conclusions}
In this paper, we have tested roles of the energy distribution of accretion disks governed by
black hole spins and accretion rates to the $R-L$ relation. Our main conclusions can be summarized in several points:
\begin{itemize}
\item New measurements of the time delays in AGN, inconsistent with the simple scaling 
relationship of \citet{bentz2013} with the monochromatic flux, favor a model where the BLR 
responds to the ionizing continuum, and both the local density and the ionization parameter 
are universal, independent from the black hole mass and Eddington ratio. Inconsistency between the results in \citet{grier2017} and \citet{bentz2013} does not seem to be related to the Eddington ratio, only slightly higher in the Grier et al. sample.

\item Since new BLR scaling is sensitive to the SED shape, the radius-luminosity relation 
is a potential tool to examine the SED from an accretion disk.  

\item If the transition to the inner hot flow is based on {\it strong ADAF principle}
 the measured delays are consistent with the model for a realistic value of the 
 viscosity parameter $\alpha$ because it does not overestimate the cold disk evaporation. 

\item If the transition to the inner hot flow is based on the disk evaporation through the electron 
conduction between the disk and the corona, low values of the magnetic pressure and very 
low values of the viscosity parameter $\alpha$ are required, so this description is less 
satisfactory than simple {\it strong ADAF principle}, and most likely implies that the inner 
cold disk formation takes place.

\item New measurements can be seen as a counter-argument against the dust-based model of the 
BLR formation since dust-based model implies a scaling of the BLR size with the monochromatic 
flux.

\item A confirmation of the new reverberation measurements for the outliers from 
the \citet{bentz2013} relationship at low/moderate Eddington ratios is strongly required.
\end{itemize}

\section*{Acknowledgements}
The project was partially supported by National Science Centre, Poland, grant No. 2017/26/A/ST9/00756 (Maestro 9). VK acknowledges Czech Science Foundation No. 17-16287S.



\bibliography{delays}

\begin{thebibliography}{}
\expandafter\ifx\csname natexlab\endcsname\relax\def\natexlab#1{#1}\fi
\providecommand{\url}[1]{\href{#1}{#1}}
\providecommand{\dodoi}[1]{doi:~\href{http://doi.org/#1}{\nolinkurl{#1}}}
\providecommand{\doeprint}[1]{\href{http://ascl.net/#1}{\nolinkurl{http://ascl.net/#1}}}
\providecommand{\doarXiv}[1]{\href{https://arxiv.org/abs/#1}{\nolinkurl{https://arxiv.org/abs/#1}}}

\bibitem[{{Abramowicz} {et~al.}(1995){Abramowicz}, {Chen}, {Kato}, {Lasota}, \&
  {Regev}}]{abramowicz1995}
{Abramowicz}, M.~A., {Chen}, X., {Kato}, S., {Lasota}, J.-P., \& {Regev}, O.
  1995, \apjl, 438, L37, \dodoi{10.1086/187709}

\bibitem[{{Ade} {et~al.}(2014){Ade}, {Aghanim}, {Armitage-Caplan}, \&
  et~al.}]{planck2013}
{Ade}, P., {Aghanim}, N., {Armitage-Caplan}, C., \& et~al. 2014, \aap, 571,
  A16, \dodoi{10.1051/0004-6361/201321591}

\bibitem[{{Baldwin} {et~al.}(1995){Baldwin}, {Ferland}, {Korista}, \&
  {Verner}}]{baldwin1995}
{Baldwin}, J., {Ferland}, G., {Korista}, K., \& {Verner}, D. 1995, \apjl, 455,
  L119, \dodoi{10.1086/309827}

\bibitem[{{Baskin} \& {Laor}(2018)}]{baskin2018}
{Baskin}, A., \& {Laor}, A. 2018, \mnras, 474, 1970,
  \dodoi{10.1093/mnras/stx2850}

\bibitem[{{Bentz} {et~al.}(2006){Bentz}, {Peterson}, {Pogge}, {Vestergaard}, \&
  {Onken}}]{bentz2006}
{Bentz}, M.~C., {Peterson}, B.~M., {Pogge}, R.~W., {Vestergaard}, M., \&
  {Onken}, C.~A. 2006, \apj, 644, 133, \dodoi{10.1086/503537}

\bibitem[{{Bentz} {et~al.}(2013){Bentz}, {Denney}, {Grier}, {Barth},
  {Peterson}, {Vestergaard}, {Bennert}, {Canalizo}, {De Rosa}, {Filippenko},
  {Gates}, {Greene}, {Li}, {Malkan}, {Pogge}, {Stern}, {Treu}, \&
  {Woo}}]{bentz2013}
{Bentz}, M.~C., {Denney}, K.~D., {Grier}, C.~J., {et~al.} 2013, \apj, 767, 149,
  \dodoi{10.1088/0004-637X/767/2/149}

\bibitem[{{Bisnovatyi-Kogan} \& {Lovelace}(1997)}]{bisnovatyi1997}
{Bisnovatyi-Kogan}, G.~S., \& {Lovelace}, R.~V.~E. 1997, \apjl, 486, L43,
  \dodoi{10.1086/310826}

\bibitem[{{Blandford} \& {Begelman}(1999)}]{blandford1999}
{Blandford}, R.~D., \& {Begelman}, M.~C. 1999, \mnras, 303, L1,
  \dodoi{10.1046/j.1365-8711.1999.02358.x}

\bibitem[{{Capellupo} {et~al.}(2015){Capellupo}, {Netzer}, {Lira},
  {Trakhtenbrot}, \& {Mej{\'{\i}}a-Restrepo}}]{capellupo2015}
{Capellupo}, D.~M., {Netzer}, H., {Lira}, P., {Trakhtenbrot}, B., \&
  {Mej{\'{\i}}a-Restrepo}, J. 2015, \mnras, 446, 3427,
  \dodoi{10.1093/mnras/stu2266}

\bibitem[{{Czerny} {et~al.}(2016){Czerny}, {Du}, {Wang}, \&
  {Karas}}]{czerny2016}
{Czerny}, B., {Du}, P., {Wang}, J.-M., \& {Karas}, V. 2016, \apj, 832, 15,
  \dodoi{10.3847/0004-637X/832/1/15}

\bibitem[{{Czerny} \& {Hryniewicz}(2011)}]{czkr2011}
{Czerny}, B., \& {Hryniewicz}, K. 2011, \aap, 525, L8,
  \dodoi{10.1051/0004-6361/201016025}

\bibitem[{{Czerny} {et~al.}(2013){Czerny}, {Hryniewicz}, {Maity},
  {Schwarzenberg-Czerny}, {{\.Z}ycki}, \& {Bilicki}}]{czerny2013}
{Czerny}, B., {Hryniewicz}, K., {Maity}, I., {et~al.} 2013, \aap, 556, A97,
  \dodoi{10.1051/0004-6361/201220832}

\bibitem[{{Czerny} {et~al.}(2011){Czerny}, {Hryniewicz}, {Niko{\l}ajuk}, \&
  {S{\c a}dowski}}]{czerny2011}
{Czerny}, B., {Hryniewicz}, K., {Niko{\l}ajuk}, M., \& {S{\c a}dowski}, A.
  2011, \mnras, 415, 2942, \dodoi{10.1111/j.1365-2966.2011.18912.x}

\bibitem[{{Czerny} {et~al.}(2004){Czerny}, {R{\'o}za{\'n}ska}, \&
  {Kuraszkiewicz}}]{czerny2004}
{Czerny}, B., {R{\'o}za{\'n}ska}, A., \& {Kuraszkiewicz}, J. 2004, \aap, 428,
  39, \dodoi{10.1051/0004-6361:20040487}

\bibitem[{{Czerny} {et~al.}(2015){Czerny}, {Modzelewska}, {Petrogalli}, {Pych},
  {Adhikari}, {{\.Z}ycki}, {Hryniewicz}, {Krupa}, {{\'S}wie{\c t}o{\'n}}, \&
  {Niko{\l}ajuk}}]{czerny2015}
{Czerny}, B., {Modzelewska}, J., {Petrogalli}, F., {et~al.} 2015, Advances in
  Space Research, 55, 1806, \dodoi{10.1016/j.asr.2015.01.004}

\bibitem[{{Czerny} {et~al.}(2017){Czerny}, {Li}, {Hryniewicz}, {Panda},
  {Wildy}, {Sniegowska}, {Wang}, {Sredzinska}, \& {Karas}}]{czerny2017}
{Czerny}, B., {Li}, Y.-R., {Hryniewicz}, K., {et~al.} 2017, \apj, 846, 154,
  \dodoi{10.3847/1538-4357/aa8810}

\bibitem[{{Du} {et~al.}(2014){Du}, {Hu}, {Lu}, {Wang}, {Qiu}, {Li}, {Bai},
  {Kaspi}, {Netzer}, {Wang}, \& {SEAMBH Collaboration}}]{du2014}
{Du}, P., {Hu}, C., {Lu}, K.-X., {et~al.} 2014, \apj, 782, 45,
  \dodoi{10.1088/0004-637X/782/1/45}

\bibitem[{{Du} {et~al.}(2015){Du}, {Hu}, {Lu}, {Huang}, {Cheng}, {Qiu}, {Li},
  {Zhang}, {Fan}, {Bai}, {Bian}, {Yuan}, {Kaspi}, {Ho}, {Netzer}, {Wang}, \&
  {SEAMBH Collaboration}}]{du2015}
---. 2015, \apj, 806, 22, \dodoi{10.1088/0004-637X/806/1/22}

\bibitem[{{Du} {et~al.}(2016){Du}, {Lu}, {Zhang}, {Huang}, {Wang}, {Hu}, {Qiu},
  {Li}, {Fan}, {Fang}, {Bai}, {Bian}, {Yuan}, {Ho}, {Wang}, \& {SEAMBH
  Collaboration}}]{du2016}
{Du}, P., {Lu}, K.-X., {Zhang}, Z.-X., {et~al.} 2016, \apj, 825, 126,
  \dodoi{10.3847/0004-637X/825/2/126}

\bibitem[{{Du} {et~al.}(2018){Du}, {Zhang}, {Wang}, {Huang}, {Zhang}, {Lu},
  {Hu}, {Li}, {Bai}, {Bian}, {Yuan}, {Ho}, {Wang}, \& {SEAMBH
  collaboration}}]{du2018}
{Du}, P., {Zhang}, Z.-X., {Wang}, K., {et~al.} 2018, \apj, 856, 6,
  \dodoi{10.3847/1538-4357/aaae6b}

\bibitem[{{Ferland} \& {Netzer}(1983)}]{ferland83}
{Ferland}, G.~J., \& {Netzer}, H. 1983, \apj, 264, 105, \dodoi{10.1086/160577}

\bibitem[{{Goad} \& {Korista}(2014)}]{goad2014}
{Goad}, M.~R., \& {Korista}, K.~T. 2014, \mnras, 444, 43,
  \dodoi{10.1093/mnras/stu1456}

\bibitem[{{Grier} {et~al.}(2008){Grier}, {Peterson}, {Bentz}, {Denney},
  {Eastman}, {Dietrich}, {Pogge}, {Prieto}, {DePoy}, {Assef}, {Atlee}, {Bird},
  {Eyler}, {Peeples}, {Siverd}, {Watson}, \& {Yee}}]{grier2008}
{Grier}, C.~J., {Peterson}, B.~M., {Bentz}, M.~C., {et~al.} 2008, \apj, 688,
  837, \dodoi{10.1086/592269}

\bibitem[{{Grier} {et~al.}(2012){Grier}, {Peterson}, {Pogge}, {Denney},
  {Bentz}, {Martini}, {Sergeev}, {Kaspi}, {Minezaki}, {Zu}, {Kochanek},
  {Siverd}, {Shappee}, {Stanek}, {Araya Salvo}, {Beatty}, {Bird}, {Bord},
  {Borman}, {Che}, {Chen}, {Cohen}, {Dietrich}, {Doroshenko}, {Drake},
  {Efimov}, {Free}, {Ginsburg}, {Henderson}, {King}, {Koshida}, {Mogren},
  {Molina}, {Mosquera}, {Nazarov}, {Okhmat}, {Pejcha}, {Rafter}, {Shields},
  {Skowron}, {Szczygiel}, {Valluri}, \& {van Saders}}]{grier2012}
{Grier}, C.~J., {Peterson}, B.~M., {Pogge}, R.~W., {et~al.} 2012, \apj, 755,
  60, \dodoi{10.1088/0004-637X/755/1/60}

\bibitem[{{Grier} {et~al.}(2017){Grier}, {Trump}, {Shen}, {Horne}, {Kinemuchi},
  {McGreer}, {Starkey}, {Brandt}, {Hall}, {Kochanek}, {Chen}, {Denney},
  {Greene}, {Ho}, {Homayouni}, {I-Hsiu Li}, {Pei}, {Peterson}, {Petitjean},
  {Schneider}, {Sun}, {AlSayyad}, {Bizyaev}, {Brinkmann}, {Brownstein},
  {Bundy}, {Dawson}, {Eftekharzadeh}, {Fernandez-Trincado}, {Gao},
  {Hutchinson}, {Jia}, {Jiang}, {Oravetz}, {Pan}, {Paris}, {Ponder}, {Peters},
  {Rogerson}, {Simmons}, {Smith}, \& {Wang}}]{grier2017}
{Grier}, C.~J., {Trump}, J.~R., {Shen}, Y., {et~al.} 2017, \apj, 851, 21,
  \dodoi{10.3847/1538-4357/aa98dc}

\bibitem[{{Grz{\c e}dzielski} {et~al.}(2017){Grz{\c e}dzielski}, {Janiuk},
  {Czerny}, \& {Wu}}]{grzedzielski2017}
{Grz{\c e}dzielski}, M., {Janiuk}, A., {Czerny}, B., \& {Wu}, Q. 2017, \aap,
  603, A110, \dodoi{10.1051/0004-6361/201629672}

\bibitem[{{Haas} {et~al.}(2011){Haas}, {Chini}, {Ramolla}, {Pozo Nu{\~n}ez},
  {Westhues}, {Watermann}, {Hoffmeister}, \& {Murphy}}]{haas2011}
{Haas}, M., {Chini}, R., {Ramolla}, M., {et~al.} 2011, \aap, 535, A73,
  \dodoi{10.1051/0004-6361/201117325}

\bibitem[{{He} {et~al.}(2018){He}, {Sun}, {Zakamska}, {Wylezalek}, {Kelly},
  {Greene}, {Rembold}, {Riffel}, \& {Riffel}}]{he2018}
{He}, Z., {Sun}, A.-L., {Zakamska}, N.~L., {et~al.} 2018, \mnras, 478, 3614,
  \dodoi{10.1093/mnras/sty1322}

\bibitem[{{Honma}(1996)}]{honma1996}
{Honma}, F. 1996, \pasj, 48, 77, \dodoi{10.1093/pasj/48.1.77}

\bibitem[{{Ichimaru}(1977)}]{ichimaru1977}
{Ichimaru}, S. 1977, \apj, 214, 840, \dodoi{10.1086/155314}

\bibitem[{{Kaspi} {et~al.}(2000){Kaspi}, {Smith}, {Netzer}, {Maoz}, {Jannuzi},
  \& {Giveon}}]{kaspi2000}
{Kaspi}, S., {Smith}, P.~S., {Netzer}, H., {et~al.} 2000, \apj, 533, 631,
  \dodoi{10.1086/308704}

\bibitem[{{Kato} \& {Nakamura}(1998)}]{kato1998}
{Kato}, S., \& {Nakamura}, K.~E. 1998, \pasj, 50, 559,
  \dodoi{10.1093/pasj/50.6.559}

\bibitem[{{Kelly} {et~al.}(2009){Kelly}, {Bechtold}, \&
  {Siemiginowska}}]{kelly2009}
{Kelly}, B.~C., {Bechtold}, J., \& {Siemiginowska}, A. 2009, \apj, 698, 895,
  \dodoi{10.1088/0004-637X/698/1/895}

\bibitem[{{Kilerci Eser} {et~al.}(2015){Kilerci Eser}, {Vestergaard},
  {Peterson}, {Denney}, \& {Bentz}}]{eser2015}
{Kilerci Eser}, E., {Vestergaard}, M., {Peterson}, B.~M., {Denney}, K.~D., \&
  {Bentz}, M.~C. 2015, \apj, 801, 8, \dodoi{10.1088/0004-637X/801/1/8}

\bibitem[{{King} {et~al.}(2014){King}, {Davis}, {Denney}, {Vestergaard}, \&
  {Watson}}]{king2014}
{King}, A.~L., {Davis}, T.~M., {Denney}, K.~D., {Vestergaard}, M., \& {Watson},
  D. 2014, \mnras, 441, 3454, \dodoi{10.1093/mnras/stu793}

\bibitem[{{King} {et~al.}(2008){King}, {Pringle}, \& {Hofmann}}]{king2008}
{King}, A.~R., {Pringle}, J.~E., \& {Hofmann}, J.~A. 2008, \mnras, 385, 1621,
  \dodoi{10.1111/j.1365-2966.2008.12943.x}

\bibitem[{{Koshida} {et~al.}(2014){Koshida}, {Minezaki}, {Yoshii}, {Kobayashi},
  {Sakata}, {Sugawara}, {Enya}, {Suganuma}, {Tomita}, {Aoki}, \&
  {Peterson}}]{koshida2014}
{Koshida}, S., {Minezaki}, T., {Yoshii}, Y., {et~al.} 2014, \apj, 788, 159,
  \dodoi{10.1088/0004-637X/788/2/159}

\bibitem[{{Krolik}(1999)}]{krolik1999}
{Krolik}, J.~H. 1999, {Active galactic nuclei : from the central black hole to
  the galactic environment}

\bibitem[{{Kubota} \& {Done}(2018)}]{kubota2018}
{Kubota}, A., \& {Done}, C. 2018, ArXiv e-prints.
\newblock \doarXiv{1804.00171}

\bibitem[{{Lawrence} \& {Elvis}(2010)}]{lawrence2010}
{Lawrence}, A., \& {Elvis}, M. 2010, \apj, 714, 561,
  \dodoi{10.1088/0004-637X/714/1/561}

\bibitem[{{Li} {et~al.}(2012){Li}, {Wang}, \& {Ho}}]{Li2012}
{Li}, Y.-R., {Wang}, J.-M., \& {Ho}, L.~C. 2012, \apj, 749, 187,
  \dodoi{10.1088/0004-637X/749/2/187}

\bibitem[{{Liu} {et~al.}(2006){Liu}, {Meyer}, \& {Meyer-Hofmeister}}]{liu2006}
{Liu}, B.~F., {Meyer}, F., \& {Meyer-Hofmeister}, E. 2006, \aap, 454, L9,
  \dodoi{10.1051/0004-6361:20065430}

\bibitem[{{Liu} {et~al.}(1999){Liu}, {Yuan}, {Meyer}, {Meyer-Hofmeister}, \&
  {Xie}}]{liu1999}
{Liu}, B.~F., {Yuan}, W., {Meyer}, F., {Meyer-Hofmeister}, E., \& {Xie}, G.~Z.
  1999, \apjl, 527, L17, \dodoi{10.1086/312383}

\bibitem[{{Liutyi}(1977)}]{liutyi1977}
{Liutyi}, V.~M. 1977, \sovast, 21, 655

\bibitem[{{Lu} {et~al.}(2016){Lu}, {Du}, {Hu}, {Li}, {Zhang}, {Wang}, {Huang},
  {Bi}, {Bai}, {Ho}, \& {Wang}}]{lu2016}
{Lu}, K.-X., {Du}, P., {Hu}, C., {et~al.} 2016, \apj, 827, 118,
  \dodoi{10.3847/0004-637X/827/2/118}

\bibitem[{{Meyer} {et~al.}(2007){Meyer}, {Liu}, \&
  {Meyer-Hofmeister}}]{meyer2007}
{Meyer}, F., {Liu}, B.~F., \& {Meyer-Hofmeister}, E. 2007, \aap, 463, 1,
  \dodoi{10.1051/0004-6361:20066203}

\bibitem[{{Meyer} \& {Meyer-Hofmeister}(1994)}]{meyer1994}
{Meyer}, F., \& {Meyer-Hofmeister}, E. 1994, \aap, 288, 175

\bibitem[{{Meyer} \& {Meyer-Hofmeister}(2002)}]{meyer02}
---. 2002, \aap, 392, L5, \dodoi{10.1051/0004-6361:20021075}

\bibitem[{{Meyer-Hofmeister} {et~al.}(2017){Meyer-Hofmeister}, {Liu}, \&
  {Qiao}}]{MeyerHofmeister2017}
{Meyer-Hofmeister}, E., {Liu}, B.~F., \& {Qiao}, E. 2017, \aap, 607, A94,
  \dodoi{10.1051/0004-6361/201731105}

\bibitem[{{Meyer-Hofmeister} \& {Meyer}(2014)}]{meyer2014}
{Meyer-Hofmeister}, E., \& {Meyer}, F. 2014, \aap, 562, A142,
  \dodoi{10.1051/0004-6361/201322423}

\bibitem[{{Narayan} \& {Yi}(1994)}]{narayan1994}
{Narayan}, R., \& {Yi}, I. 1994, \apjl, 428, L13, \dodoi{10.1086/187381}

\bibitem[{{Novikov} \& {Thorne}(1973)}]{novikov1973}
{Novikov}, I.~D., \& {Thorne}, K.~S. 1973, in Black Holes (Les Astres Occlus),
  ed. C.~{Dewitt} \& B.~S. {Dewitt}, 343--450

\bibitem[{{Pancoast} {et~al.}(2014){Pancoast}, {Brewer}, {Treu}, {Park},
  {Barth}, {Bentz}, \& {Woo}}]{pancoast2014}
{Pancoast}, A., {Brewer}, B.~J., {Treu}, T., {et~al.} 2014, \mnras, 445, 3073,
  \dodoi{10.1093/mnras/stu1419}

\bibitem[{{Peterson} {et~al.}(2004){Peterson}, {Ferrarese}, {Gilbert}, {Kaspi},
  {Malkan}, {Maoz}, {Merritt}, {Netzer}, {Onken}, {Pogge}, {Vestergaard}, \&
  {Wandel}}]{peterson2004}
{Peterson}, B.~M., {Ferrarese}, L., {Gilbert}, K.~M., {et~al.} 2004, \apj, 613,
  682, \dodoi{10.1086/423269}

\bibitem[{{Richards} {et~al.}(2006){Richards}, {Lacy}, {Storrie-Lombardi},
  {Hall}, {Gallagher}, {Hines}, {Fan}, {Papovich}, {Vanden Berk}, {Trammell},
  {Vestergaard}, {York}, {Jester}, {Anderson}, {Budav{\'a}ri}, \&
  {Szalay}}]{richards2006}
{Richards}, G.~T., {Lacy}, M., {Storrie-Lombardi}, L.~J., {et~al.} 2006, \apjs,
  166, 470, \dodoi{10.1086/506525}

\bibitem[{{R{\'o}{\.z}a{\'n}ska} \& {Czerny}(2000{\natexlab{a}})}]{rozanska00}
{R{\'o}{\.z}a{\'n}ska}, A., \& {Czerny}, B. 2000{\natexlab{a}}, \mnras, 316,
  473, \dodoi{10.1046/j.1365-8711.2000.03429.x}

\bibitem[{{R{\'o}{\.z}a{\'n}ska} \&
  {Czerny}(2000{\natexlab{b}})}]{rozanska2000}
---. 2000{\natexlab{b}}, \aap, 360, 1170

\bibitem[{{Shen} {et~al.}(2011){Shen}, {Richards}, {Strauss}, {Hall},
  {Schneider}, {Snedden}, {Bizyaev}, {Brewington}, {Malanushenko},
  {Malanushenko}, {Oravetz}, {Pan}, \& {Simmons}}]{shen2011}
{Shen}, Y., {Richards}, G.~T., {Strauss}, M.~A., {et~al.} 2011, \apjs, 194, 45,
  \dodoi{10.1088/0067-0049/194/2/45}

\bibitem[{{Shen} {et~al.}(2015){Shen}, {Greene}, {Ho}, {Brandt}, {Denney},
  {Horne}, {Jiang}, {Kochanek}, {McGreer}, {Merloni}, {Peterson}, {Petitjean},
  {Schneider}, {Schulze}, {Strauss}, {Tao}, {Trump}, {Pan}, \&
  {Bizyaev}}]{shen2015}
{Shen}, Y., {Greene}, J.~E., {Ho}, L.~C., {et~al.} 2015, \apj, 805, 96,
  \dodoi{10.1088/0004-637X/805/2/96}

\bibitem[{{Siemiginowska} \& {Czerny}(1989)}]{siemiginowska1989}
{Siemiginowska}, A., \& {Czerny}, B. 1989, \mnras, 239, 289,
  \dodoi{10.1093/mnras/239.1.289}

\bibitem[{{Sironi} \& {Narayan}(2015)}]{sironi2015}
{Sironi}, L., \& {Narayan}, R. 2015, \apj, 800, 88,
  \dodoi{10.1088/0004-637X/800/2/88}

\bibitem[{{Starling} {et~al.}(2004){Starling}, {Siemiginowska}, {Uttley}, \&
  {Soria}}]{starling2004}
{Starling}, R.~L.~C., {Siemiginowska}, A., {Uttley}, P., \& {Soria}, R. 2004,
  \mnras, 347, 67, \dodoi{10.1111/j.1365-2966.2004.07167.x}

\bibitem[{{Taam} {et~al.}(2018){Taam}, {Qiao}, {Liu}, \&
  {Meyer-Hofmeister}}]{taam2018}
{Taam}, R.~E., {Qiao}, E., {Liu}, B.~F., \& {Meyer-Hofmeister}, E. 2018, \apj,
  860, 166, \dodoi{10.3847/1538-4357/aac50d}

\bibitem[{{Tovmassian}(2001)}]{tovmassian2001}
{Tovmassian}, H.~M. 2001, Astronomische Nachrichten, 322, 87,
  \dodoi{10.1002/1521-3994(200106)322:2<87::AID-ASNA87>3.0.CO;2-S}

\bibitem[{{Trippe}(2015)}]{trippe2015}
{Trippe}, S. 2015, Journal of Korean Astronomical Society, 48, 203,
  \dodoi{10.5303/JKAS.2015.48.3.203}

\bibitem[{{Vestergaard} \& {Peterson}(2006)}]{vestergaard2006}
{Vestergaard}, M., \& {Peterson}, B.~M. 2006, \apj, 641, 689,
  \dodoi{10.1086/500572}

\bibitem[{{Volonteri} {et~al.}(2013){Volonteri}, {Sikora}, {Lasota}, \&
  {Merloni}}]{volonteri2013}
{Volonteri}, M., {Sikora}, M., {Lasota}, J.-P., \& {Merloni}, A. 2013, \apj,
  775, 94, \dodoi{10.1088/0004-637X/775/2/94}

\bibitem[{{Wandel} {et~al.}(1999){Wandel}, {Peterson}, \&
  {Malkan}}]{wandel1999}
{Wandel}, A., {Peterson}, B.~M., \& {Malkan}, M.~A. 1999, \apj, 526, 579,
  \dodoi{10.1086/308017}

\bibitem[{{Wang} {et~al.}(2016){Wang}, {Du}, {Hu}, {Bai}, {Wang}, {Yi}, {Wang},
  {Zhang}, {Xin}, {Lun}, {Chang}, \& {Fan}}]{wang2016}
{Wang}, F., {Du}, P., {Hu}, C., {et~al.} 2016, \apj, 824, 149,
  \dodoi{10.3847/0004-637X/824/2/149}

\bibitem[{{Wang} {et~al.}(2012){Wang}, {Du}, {Baldwin}, {Ge}, {Hu}, \&
  {Ferland}}]{wang2012}
{Wang}, J.-M., {Du}, P., {Baldwin}, J.~A., {et~al.} 2012, \apj, 746, 137,
  \dodoi{10.1088/0004-637X/746/2/137}

\bibitem[{{Wang} {et~al.}(2017){Wang}, {Du}, {Brotherton}, {Hu}, {Songsheng},
  {Li}, {Shi}, \& {Zhang}}]{wang2017}
{Wang}, J.-M., {Du}, P., {Brotherton}, M.~S., {et~al.} 2017, Nature Astronomy,
  1, 775, \dodoi{10.1038/s41550-017-0264-4}

\bibitem[{{Wang} {et~al.}(2014{\natexlab{a}}){Wang}, {Du}, {Li}, {Ho}, {Hu}, \&
  {Bai}}]{wang2014}
{Wang}, J.-M., {Du}, P., {Li}, Y.-R., {et~al.} 2014{\natexlab{a}}, \apjl, 792,
  L13, \dodoi{10.1088/2041-8205/792/1/L13}

\bibitem[{{Wang} {et~al.}(2014{\natexlab{b}}){Wang}, {Qiu}, {Du}, \&
  {Ho}}]{wang_shield2014}
{Wang}, J.-M., {Qiu}, J., {Du}, P., \& {Ho}, L.~C. 2014{\natexlab{b}}, \apj,
  797, 65, \dodoi{10.1088/0004-637X/797/1/65}

\bibitem[{{Wang} {et~al.}(2009){Wang}, {Hu}, {Li}, {Chen}, {King}, {Marconi},
  {Ho}, {Yan}, {Staubert}, \& {Zhang}}]{wang2009}
{Wang}, J.-M., {Hu}, C., {Li}, Y.-R., {et~al.} 2009, \apjl, 697, L141,
  \dodoi{10.1088/0004-637X/697/2/L141}

\bibitem[{{Wang} {et~al.}(2011){Wang}, {Ge}, {Hu}, {Baldwin}, {Li}, {Ferland},
  {Xiang}, {Yan}, \& {Zhang}}]{wang2011}
{Wang}, J.-M., {Ge}, J.-Q., {Hu}, C., {et~al.} 2011, \apj, 739, 3,
  \dodoi{10.1088/0004-637X/739/1/3}

\bibitem[{{Watson} {et~al.}(2011){Watson}, {Denney}, {Vestergaard}, \&
  {Davis}}]{watson2011}
{Watson}, D., {Denney}, K.~D., {Vestergaard}, M., \& {Davis}, T.~M. 2011,
  \apjl, 740, L49, \dodoi{10.1088/2041-8205/740/2/L49}

\bibitem[{{Yuan} \& {Narayan}(2014)}]{yuan2014}
{Yuan}, F., \& {Narayan}, R. 2014, \araa, 52, 529,
  \dodoi{10.1146/annurev-astro-082812-141003}

\end{thebibliography}
\bibliographystyle{aasjournal}

\startlongtable
\begin{deluxetable}{lrrrrl}
\tablecaption{Time delays \label{tab:objects}}
\tablehead{
\colhead{Name}                & 
\colhead{$\log L_{5100}$}     & 
\colhead{$R_{\rm H\beta}$}    &
\colhead{$\log M_{BH}$}       &
\colhead{$L/L_{Edd}$}          &
\colhead{Ref}                    \\
\colhead{}                    & 
\colhead{($\rm erg/s$)}       & 
\colhead{(ldt)}               &
\colhead{$M_{\odot}$}         
}
\startdata
SDSS J140518    &    44.33    &    $41.6^{+14.8}_{-8.3}$    &      7.74   &   0.288 &  1 \\ 
SDSS J140759    &    43.58    &    $16.3^{+13.1}_{-6.6}$    &      7.67   &   0.059 &   \\  
SDSS J140812    &    43.15    &    $10.5^{+1.0}_{-2.2}$     &      7.26   &   0.058 &   \\  
SDSS J140904    &    44.15    &    $11.6^{+8.6}_{-4.6}$     &      8.45   &   0.036 &   \\  
SDSS J141004    &    44.22    &    $53.5^{+4.2}_{-4.0}$     &      8.32   &   0.058 &   \\  
SDSS J141018    &    43.58    &    $16.2^{+2.9}_{-4.5}$     &      7.65   &   0.063 &   \\  
SDSS J141031    &    44.02    &    $35.8^{+1.1}_{-10.3}$    &      7.91   &   0.094 &   \\  
SDSS J141041    &    43.82    &    $21.9^{+4.2}_{-2.4}$     &      7.85   &   0.070 &   \\  
SDSS J141112    &    44.12    &    $20.4^{+2.5}_{-2.0}$     &      7.41   &   0.375 &   \\  
SDSS J141115    &    44.31    &    $49.1^{+11.1}_{-2.0}$    &      7.94   &   0.174 &   \\  
SDSS J141123    &    44.13    &    $13.0^{+1.4}_{-0.8}$     &      7.38   &   0.411 &   \\  
SDSS J141135    &    44.04    &    $17.6^{+8.6}_{-7.4}$     &      7.56   &   0.224 &   \\  
SDSS J141147    &    44.02    &    $6.4^{+1.5}_{-1.4}$      &      6.95   &   0.865 &   \\  
SDSS J141214    &    44.40    &    $21.4^{+4.2}_{-6.4}$     &      7.26   &   1.019 &   \\  
SDSS J141314    &    44.52    &    $43.9^{+4.9}_{-4.3}$     &      9.21   &   0.015 &   \\  
SDSS J141318    &    43.94    &    $20.0^{+1.1}_{-3.0}$     &      7.51   &   0.200 &   \\  
SDSS J141324    &    43.94    &    $25.5^{+10.9}_{-5.8}$    &      8.92   &   0.008 &   \\  
SDSS J141417    &    43.40    &    $15.6^{+3.2}_{-5.1}$     &      8.03   &   0.017 &   \\  
SDSS J141532    &    44.14    &    $26.5^{+9.9}_{-8.8}$     &      7.23   &   0.591 &   \\  
SDSS J141606    &    44.80    &    $32.0^{+11.6}_{-15.5}$   &      9.07   &   0.040 &   \\  
SDSS J141625    &    43.96    &    $15.1^{+3.2}_{-4.6}$     &      7.58   &   0.178 &   \\  
SDSS J141645.15 &    43.21    &    $5.0^{+1.5}_{-1.4}$      &      7.93   &   0.014 &   \\  
SDSS J141645.58 &    43.68    &    $8.5^{+2.5}_{-1.4}$      &      6.90   &   0.438 &   \\  
SDSS J141706    &    44.19    &    $10.4^{+6.3}_{-3.0}$     &      6.70   &   2.258 &   \\  
SDSS J141712    &    43.21    &    $12.5^{+1.8}_{-2.6}$     &      8.99   &   0.001 &   \\  
SDSS J141724    &    43.99    &    $10.1^{+12.5}_{-2.7}$    &      7.57   &   0.195 &   \\  
SDSS J141729    &    43.29    &    $5.5^{+5.7}_{-2.1}$      &      8.28   &   0.008 &   \\  
SDSS J141856    &    45.38    &    $15.8^{+6.0}_{-1.9}$     &      8.90   &   0.224 &   \\  
SDSS J141859    &    44.91    &    $20.4^{+5.6}_{-7.0}$     &      8.05   &   0.534 &   \\  
SDSS J141923    &    43.12    &    $11.8^{+0.7}_{-1.5}$     &      7.18   &   0.065 &   \\  
SDSS J141941    &    44.52    &    $30.4^{+3.9}_{-8.3}$     &      7.60   &   0.612 &   \\  
SDSS J141952    &    44.25    &    $32.9^{+5.6}_{-5.1}$     &      9.23   &   0.008 &   \\  
SDSS J141955    &    43.40    &    $10.7^{+5.6}_{-4.4}$     &      7.69   &   0.037 &   \\  
SDSS J142010    &    44.09    &    $12.8^{+5.7}_{-4.5}$     &      8.64   &   0.021 &   \\  
SDSS J142023    &    44.22    &    $8.5^{+3.2}_{-3.9}$      &      8.58   &   0.032 &   \\  
SDSS J142038    &    43.46    &    $25.2^{+4.7}_{-5.7}$     &      7.67   &   0.045 &   \\  
SDSS J142039    &    44.14    &    $20.7^{+0.9}_{-3.0}$     &      7.57   &   0.275 &   \\  
SDSS J142043    &    43.40    &    $5.9^{+0.4}_{-0.6}$      &      7.20   &   0.115 &   \\  
SDSS J142049    &    44.45    &    $46.0^{+9.5}_{-9.5}$     &      9.00   &   0.020 &   \\  
SDSS J142052    &    45.06    &    $11.9^{+1.3}_{-1.0}$     &      8.73   &   0.157 &   \\  
SDSS J142103    &    43.64    &    $75.2^{+3.2}_{-3.3}$     &      7.89   &   0.041 &   \\  
SDSS J142112    &    44.31    &    $14.2^{+3.7}_{-3.0}$     &      8.22   &   0.092 &   \\  
SDSS J142135    &    43.47    &    $3.9^{+0.9}_{-0.9}$      &      6.60   &   0.548 &   \\  
SDSS J142417    &    44.09    &    $36.3^{+4.5}_{-5.5}$     &      7.70   &   0.180 &   \\ \hline
Mrk335          &    43.76    &    $14.0^{+4.6}_{-3.4}$     &      6.93   &   0.501 &  2\\  
Mrk142          &    43.59    &    $6.4^{+7.3}_{-3.4}$      &      6.47   &   0.983 &   \\  
IRASF12397      &    44.23    &    $9.7^{+5.5}_{-1.8}$      &      6.79   &   2.023 &   \\  
Mrk486          &    43.69    &    $23.7^{+7.5}_{-2.7}$     &      7.24   &   0.208 &   \\  
Mrk382          &    43.12    &    $7.5^{+2.9}_{-2.0}$      &      6.50   &   0.312 &   \\  
IRAS04416       &    44.47    &    $13.3^{+13.9}_{-1.4}$    &      6.78   &   3.577 &   \\  
MCG06           &    42.67    &    $24.0^{+8.4}_{-4.8}$     &      6.92   &   0.042 &  \\   
Mrk493          &    43.11    &    $11.6^{+1.2}_{-2.6}$     &      6.14   &   0.694 &   \\  
Mrk1044         &    43.10    &    $10.5^{+3.3}_{-2.7}$     &      6.45   &   0.322 &   \\  
SDSS J074352    &    45.37    &    $43.9^{+5.2}_{-4.2}$     &      7.93   &   2.028 &   \\  
SDSS J075051    &    45.33    &    $66.6^{+18.7}_{-9.9}$    &      7.67   &   3.359 &   \\  
SDSS J075101    &    44.18    &    $30.4^{+7.3}_{-5.8}$     &      7.18   &   0.733 &   \\  
SDSS J075949    &    44.20    &    $43.9^{+33.1}_{-19.0}$   &      7.44   &   0.415 &   \\  
SDSS J080101    &    44.27    &    $8.3^{+9.7}_{-2.7}$      &      6.78   &   2.257 &   \\  
SDSS J080131    &    43.97    &    $11.5^{+7.5}_{-3.7}$     &      6.51   &   2.121 &   \\  
SDSS J081441    &    43.96    &    $25.3^{+10.4}_{-7.5}$    &      7.18   &   0.446 &   \\  
SDSS J081456    &    43.99    &    $24.3^{+7.7}_{-16.4}$    &      7.44   &   0.259 &   \\  
SDSS J083553    &    44.44    &    $12.4^{+5.4}_{-5.4}$     &      6.87   &   2.722 &   \\  
SDSS J084533    &    44.53    &    $18.1^{+6.0}_{-4.7}$     &      6.76   &   4.264 &   \\  
SDSS J085946    &    44.41    &    $34.8^{+9.2}_{-26.3}$    &      7.30   &   0.952 &   \\  
SDSS J093302    &    44.31    &    $19.0^{+3.8}_{-4.3}$     &      7.08   &   1.245 &   \\  
SDSS J093922    &    44.07    &    $11.9^{+2.1}_{-6.3}$     &      6.53   &   2.564 &   \\  
SDSS J100402    &    45.52    &    $32.2^{+43.5}_{-4.2}$    &      7.44   &   8.802 &   \\  
SDSS J101000    &    44.76    &    $27.7^{+23.5}_{-7.6}$    &      7.46   &   1.456 &   \\  
SDSS J102339    &    44.09    &    $24.9^{+19.8}_{-3.9}$    &      7.16   &   0.618 &   \\ \hline
NGC 5548        &    43.21    &    $7.2^{+1.3}_{-0.3}$      &      7.94   &   0.014 & 3  \\ 
1H 0323+342     &    43.88    &    $14.8^{+3.9}_{-2.7}$     &      7.53   &   0.164 & 4 \\ \hline
PG0026+129      &    44.97    &    $111.0^{+24.1}_{-28.3}$  &      8.15   &   0.489 & 5 \\  
PG0052+251      &    44.81    &    $89.8^{+24.5}_{-24.1}$   &      8.64   &   0.107 &   \\  
Fairall9        &    43.98    &    $17.4^{+3.2}_{-4.3}$     &      8.09   &   0.058 &   \\  
Mrk590          &    43.50    &    $25.6^{+6.5}_{-5.3}$     &      7.55   &   0.065 &   \\  
3C120           &    44.00    &    $26.2^{+8.7}_{-6.6}$     &      7.79   &   0.122 &   \\  
Ark120          &    43.87    &    $39.5^{+8.5}_{-7.8}$     &      8.47   &   0.018 &   \\  
Mrk79           &    43.68    &    $15.6^{+5.1}_{-4.9}$     &      7.84   &   0.050 &   \\  
PG0804+761      &    44.91    &    $146.9^{+18.8}_{-18.9}$  &      8.43   &   0.223 &   \\  
Mrk110          &    43.66    &    $25.6^{+8.9}_{-7.2}$     &      7.10   &   0.265 &   \\  
PG0953+414      &    45.19    &    $150.1^{+21.6}_{-22.6}$  &      8.44   &   0.408 &   \\  
NGC3227         &    42.24    &    $3.8^{+0.8}_{-0.8}$      &      7.09   &   0.010 &   \\  
NGC3516         &    42.79    &    $11.7^{+1.0}_{-1.5}$     &      7.82   &   0.007 &   \\  
SBS1116+583A    &    42.14    &    $2.3^{+0.6}_{-0.5}$      &      6.78   &   0.017 &   \\  
Arp151          &    42.55    &    $4.0^{+0.5}_{-0.7}$      &      6.87   &   0.035 &   \\  
NGC3783         &    42.56    &    $10.2^{+3.3}_{-2.3}$     &      7.45   &   0.009 &   \\  
Mrk1310         &    42.29    &    $3.7^{+0.6}_{-0.6}$      &      6.62   &   0.035 &   \\  
NGC4051         &    41.90    &    $2.1^{+0.9}_{-0.7}$      &      5.72   &   0.110 &   \\  
NGC4151         &    42.09    &    $6.6^{+1.1}_{-0.8}$      &      7.72   &   0.002 &   \\  
Mrk202          &    42.26    &    $3.0^{+1.7}_{-1.1}$      &      6.11   &   0.104 &   \\  
NGC4253         &    42.57    &    $6.2^{+1.6}_{-1.2}$      &      6.49   &   0.088 &   \\  
PG1226+023      &    45.96    &    $306.8^{+68.5}_{-90.9}$  &      8.87   &   0.918 &   \\  
PG1229+204      &    43.70    &    $37.8^{+27.6}_{-15.3}$   &      8.03   &   0.034 &   \\  
NGC4593         &    42.62    &    $4.0^{+0.8}_{-0.7}$      &      7.26   &   0.017 &   \\  
NGC4748         &    42.56    &    $5.5^{+1.6}_{-2.2}$      &      6.61   &   0.064 &   \\  
PG1307+085      &    44.85    &    $105.6^{+36.0}_{-46.6}$  &      8.72   &   0.098 &   \\  
Mrk279          &    43.71    &    $16.7^{+3.9}_{-3.9}$     &      7.97   &   0.040 &   \\  
PG1411+442      &    44.56    &    $124.3^{+61.0}_{-61.7}$  &      8.28   &   0.141 &   \\  
NGC5548         &    43.29    &    $17.6^{+6.4}_{-4.7}$     &      7.94   &   0.014 &   \\  
PG1426+015      &    44.63    &    $95.0^{+29.9}_{-37.1}$   &      8.97   &   0.033 &   \\  
Mrk817          &    43.74    &    $19.9^{+9.9}_{-6.7}$     &      7.99   &   0.042 &   \\  
Mrk290          &    43.17    &    $8.7^{+1.2}_{-1.0}$      &      7.55   &   0.031 &   \\  
PG1613+658      &    44.77    &    $40.1^{+15.0}_{-15.2}$   &      8.81   &   0.068 &   \\  
PG1617+175      &    44.39    &    $71.5^{+29.6}_{-33.7}$   &      8.79   &   0.029 &   \\  
PG1700+518      &    45.59    &    $251.8^{+45.9}_{-38.8}$  &      8.40   &   1.137 &   \\  
3C390.3         &    44.43    &    $44.5^{+27.6}_{-17.0}$   &      9.18   &   0.013 &   \\  
NGC6814         &    42.12    &    $6.6^{+0.9}_{-0.9}$      &      7.16   &   0.007 &   \\  
Mrk509          &    44.19    &    $79.6^{+6.1}_{-5.4}$     &      8.15   &   0.081 &   \\  
PG2130+099      &    44.20    &    $9.6^{+1.2}_{-1.2}$      &      7.05   &   1.043 &   \\  
NGC7469         &    43.51    &    $10.8^{+3.4}_{-1.3}$     &      7.60   &   0.059 &   \\  
PG1211+143      &    44.73    &    $93.8^{+25.6}_{-42.1}$   &      7.87   &   0.530 &   \\  
PG0844+349      &    44.22    &    $32.3^{+13.7}_{-13.4}$   &      7.66   &   0.265 &   \\  
NGC5273         &    41.54    &    $2.2^{+1.2}_{-1.6}$      &      7.14   &   0.002 &   \\  
KA1858+4850     &    43.43    &    $13.5^{+2.0}_{-2.3}$     &      6.94   &   0.225 &   \\  
Mrk1511         &    43.16    &    $5.7^{+0.9}_{-0.8}$      &      7.29   &   0.055 &   \\  
MCG6-30-15      &    41.64    &    $5.7^{+1.8}_{-1.7}$      &      6.63   &   0.008 &   \\  
UGC06728        &    41.86    &    $1.4^{+0.7}_{-0.8}$      &      5.87   &   0.073 &   \\  
MCG+08-11-011   &    43.33    &    $15.7^{+0.5}_{-0.5}$     &      7.72   &   0.030 &   \\  
NGC2617         &    42.67    &    $4.3^{+1.1}_{-1.4}$      &      7.74   &   0.006 &   \\  
3C382           &    43.84    &    $40.5^{+8.0}_{-3.7}$     &      8.67   &   0.011 &   \\  
Mrk374          &    43.77    &    $14.8^{+5.8}_{-3.3}$     &      7.86   &   0.061 &   \\ \hline
\enddata
\tablecomments{Ref. 
1: \cite{grier2017}, 
2: \cite{du2014,du2015,du2016,du2018},
3: \cite{lu2016},
4: \cite{wang2016},
5: \cite{bentz2013}}
\end{deluxetable}

\end{document}